\documentclass{egpubl}
\electronicVersion 

\usepackage[pdftex]{graphicx} \pdfcompresslevel=9

\PrintedOrElectronic

\usepackage{t1enc,dfadobe}
\usepackage{egweblnk}
\usepackage{cite}

\title{Primal-Dual Optimization for Fluids}
\author[T. Inglis, M.-L. Eckert, J. Gregson \& Nils Thuerey]
{T. Inglis$^{1 \ast}$,
	M.-L. Eckert$^{1 \ast}$,
	J. Gregson$^{2}$,
	N. Thuerey$^{1}$
	\\
	$^1$Technical University of Munich\quad
	$^2$University of British Columbia
	\\
	$^{\ast}$These authors contributed equally to this work}
	
\usepackage{algorithm}
\usepackage[noend]{algpseudocode}
\usepackage{amsmath}
\usepackage{multirow}

\newcommand{\myrefalg}[1]{Algorithm~(\ref{#1})}
\newcommand{\myrefeq}[1]{Eq.~(\ref{#1})}
\newcommand{\myreffig}[1]{Figure~\ref{#1}}
\newcommand{\myreftab}[1]{Table~\ref{#1}}
\newcommand{\myrefsec}[1]{Section~\ref{#1}}
\newcommand{\myrefsecs}[2]{Sections~\ref{#1} and \ref{#2}}
\newcommand{\myrefapp}[1]{Appendix~\ref{#1}}

\usepackage{subcaption}
\usepackage{xspace}

\DeclareMathOperator*{\argmin}{arg\,min}
\newcommand{\prox}{\mathrm{\mathbf{prox}}} 
\renewcommand{\vec}[1]{\mathbf{#1}}
\newcommand{\norm}[1]{\left\lVert#1\right\rVert} 

\newcommand{\naive}{na\"{i}ve\xspace}
\newcommand{\Naive}{Na\"{i}ve\xspace}
\newcommand{\naively}{na\"{i}vely\xspace}

\newcommand{\x}{\vec{x}} 
\newcommand{\z}{\vec{z}} 
\newcommand{\y}{\vec{y}} 
\newcommand{\q}{\vec{q}} 
\renewcommand{\u}{\vec{u}}

\newcommand{\uc}{\vec{u}_c} 
\newcommand{\ut}{\vec{u}_t} 
\newcommand{\gv}{\vec{\xi}} 
\renewcommand{\r}{\vec{r}} 
\newcommand{\blurR}{\beta} 
\newcommand{\blurM}{G} 
\newcommand{\SVW}{W} 
\newcommand{\half}{\tfrac{1}{2}} 
\newcommand{\nDim}{n_\mathrm{dim}}
\newcommand{\ratio}{r} 
\renewcommand{\div}{\mathrm{DIV}}
\newcommand{\cdiv}{C_{\div}}

\newcommand{\abs}{\mathrm{abs}}
\newcommand{\rel}{\mathrm{rel}}

\newcommand{\epsAbs}{\epsilon_{\abs}}
\newcommand{\epsRel}{\epsilon_{\rel}}

\newcommand{\epsCG}{\epsilon_{\mathrm{CG}}}
\newcommand{\dif}{\epsilon_{\mathrm{dif}}}
\newcommand{\tmp}{\epsilon_{\mathrm{tmp}}}

\newcommand{\PiB}{\Pi}
\newcommand{\PiDiv}{\Pi_{\mathrm{DIV}}}

\newcommand{\FSn}{S_{\mathrm{nsep}}}

\newlength{\punctuationfootlength}
\newcommand{\punctuationfootnote}[2]{#2\settowidth{\punctuationfootlength}%
{#2}\hspace{-\punctuationfootlength}\footnote{#1}}

\newcommand{\new}{\mathrm{new}}

\begin{document}

\teaser{
   \includegraphics[width=0.99\textwidth]{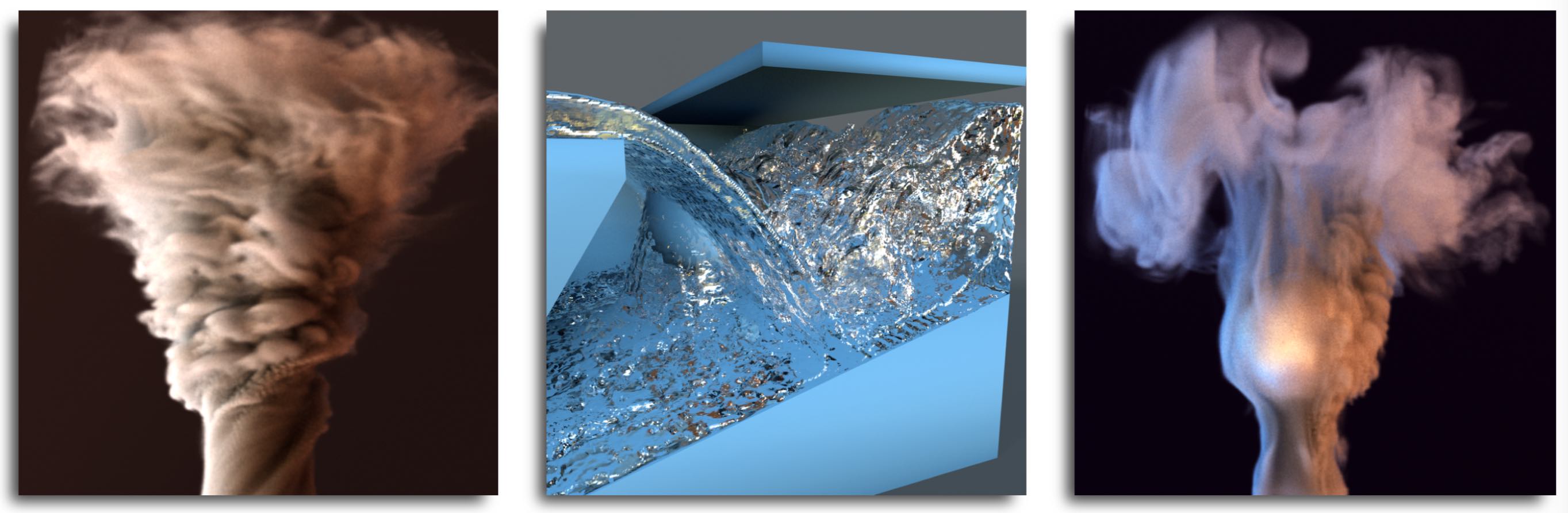}
   \caption{ \label{fig:teaser}
	Our modular Primal-Dual optimization method can be applied to fluid guiding (left, right) and to simulate liquids with separating boundary conditions (center).}
}

\maketitle


\begin{abstract}
	
	We apply a novel optimization scheme from the image processing and machine learning areas, a fast Primal-Dual method, to achieve controllable and realistic fluid simulations. 
	While our method is generally applicable to many problems in fluid simulations, we focus on the two topics of fluid guiding and separating solid-wall boundary conditions. 
	Each problem is posed as an optimization problem and solved using our method, which contains acceleration schemes tailored to each problem.
	In fluid guiding, we are interested in partially guiding fluid motion to exert control while preserving fluid characteristics. 
	With our method, we achieve explicit control over both large-scale motions and small-scale details which is valuable for many applications, 
	such as level-of-detail adjustment (after running the coarse simulation), spatially varying guiding strength, domain modification, and resimulation with different fluid parameters.
	For the separating solid-wall boundary conditions problem, our method effectively eliminates unrealistic artifacts of fluid crawling up solid walls and sticking to ceilings, requiring few changes to existing implementations. 
	We demonstrate the fast convergence of our Primal-Dual method with a variety of test cases for both model problems.
	
\end{abstract}

\section{Introduction}

Advances in fluid simulation have had a tremendous effect in engineering and
graphics.  Since fluids play an important role in our everyday lives, smoke and
liquid simulations now routinely appear as regular elements in feature films,
television and commercials.  While engineering applications are primarily
concerned with accuracy, the focus in graphics lies in simulating realistic
behavior that can be controlled to tell a story.  Therefore, it is important to
provide controllable and visually plausible fluid solvers.  Visual plausibility
is crucial since people easily recognize unrealistic behavior due to
their daily interactions with a wide range of fluid phenomena, e.g., pouring a
glass of water, boiling a kettle, or driving past a chimney.

However, it is a recurring challenge to simultaneously achieve controllable and
realistic behavior.  Fluids are usually chaotic at human scales; miniscule
perturbations regularly trigger large-scale behavior.  It is almost
impossible for artists to add small-scale details to a low-resolution
simulation without changing the large-scale motion.  Even more challenging 
is to modify the domain or rerun a simulation with different fluid
parameters without affecting the large-scale motion. 
These problems can be mitigated by guiding the velocity within an optimization framework. 
We realize guiding by constraining the velocity at each time step to
be arbitrarily close to the current and the target velocity. At the same time,
we ensure that the resulting motion is divergence-free, and we allow for
spatially varying guiding parameters.
This framework also benefits animators in the special effects industry who
may wish to create fictitious but visually plausible fluid motion to increase
entertainment value.

Common fluid solvers typically feature boundary conditions (BCs) that lead to fluid
unnaturally crawling up walls and sticking to ceilings, diminishing its
visual plausibility.  The culprit is the solid-wall BC that enforces a normal velocity of zero at obstacle
walls, thus preventing fluids from separating.
While it is physically correct to leave a thin film of fluid on the wall, its thickness in simulations is on the order of the discretization and usually far too big for realistic animations.
One proposed remedy by Batty et al.~\cite{Batty2007} is to integrate
inequality constraints into the pressure solve to allow for positive normal
velocities at solid walls. Unfortunately, this greatly increases the complexity of
the pressure solve of a fluid solver, and typically requires Quadratic
Programming solvers. 
Our goal is to implement flexible separating BCs, while reusing the popular preconditioned conjugate gradient method. 
First, we classify boundary cells into separating and non-separating cells. 
We then enforce solid-wall BCs for the velocity at non-separating cells while ensuring zero divergence.

To guide simulations and realize separating BCs, we take inspiration from convex optimization. 
Gregson et al.~\cite{GregsonCapture} already established a connection between fluid pressure correction and convex optimization. 
This connection allows us to impose all required physical constraints on the velocity for both guiding and separating BCs via an efficient alternating minimization algorithm 
that exploits problem structure, removing the need for a monolithic and slow framework.
We introduce a novel optimization scheme from the computer vision area, the \emph{fast}, or \emph{first-order
	Primal-Dual} (PD) method proposed by Pock et al.~\cite{pock2009algorithm}. It features an optimal convergence rate for non-smooth convex problems.
Note that a whole class of primal-dual methods exists, but we will hereafter use the abbreviation {\em PD} to refer to this particular instance,
which we will focus on due to its fast convergence properties.

Specifically, our contributions are
\begin{itemize}
	\item the introduction of a modular convex optimization approach for fluids based on the PD method;
	\item a general method for handling the difficult problem of fluid guiding that involves spatially varying operators;
	\item a fast method to approximately invert the linear system for flow guiding in arbitrary domains; and
	\item a novel, practical way to handle separating BCs for liquids.
\end{itemize}

\vspace{-0.1cm}
\section{Previous Work}

Fluid simulation has a long history in computer graphics \cite{cg:kass:1990}.  One of the most widely used methods is the 
stable fluid solver \cite{stam1999}, for which many extensions have been proposed over the years, e.g.,
the grid-particle hybrid {\em FLIP} (Fluid Implicit Particle) method \cite{Zhu2005}, which is particularly popular for liquid simulations. A good overview of these methods is given in Bridson's book~\cite{Bridson200809}.
A key component of incompressible fluid simulation is ensuring that the
time-evolved velocity is divergence-free (i.e., mass-preserving),
which is typically implemented via Chorin-like
pressure-projections, e.g. ~\cite{foster2001practical} or ~\cite{Batty2007} for variational approaches.

\paragraph*{Convex Optimization}
has become a powerful component of many computer vision algorithms.
Even before, the works of Boyd et al.~\cite{Boyd:ADMM} 
have laid the foundations for many popular optimization algorithms.
Closer to computer graphics, Heide et al.~\cite{Heide2013}
have proposed using the PD method to correct the deficiencies of simple lenses.
Recently, Heide et al.~\cite{pd2016} have solved the difficult task of choosing the optimal image priors and optimization algorithms for image processing tasks such as deconvolution, denoising or inpainting by applying PD as well.
Iterated Orthogonal Projection (IOP), the algorithm proposed by Molemaker et al.~\cite{MolemakerLow}, is a specialized method in the convex optimization area requiring all operators to be orthogonal projections.
Approaches such as position-based fluids (PBF)~\cite{posBasedFluids2013} also can be seen as an iterative application of a set of constraint projections. As the constraints are directly applied to the degrees of freedom, PBF is more closely related to IOP than PD, which achieves improved convergence by projecting the convex conjugate.

We extend upon the ideas presented by Gregson et al.~\cite{GregsonCapture}, where the method Alternating Direction Methods of Multipliers (ADMM) is applied, by introducing a more efficient splitting algorithm for flow guiding and BCs. 
Very recently, Narain et al.~\cite{narain2016admm} animated deformable objects with an ADMM algorithm combining a fast and robust nonlinear elasticity model with hard constraints. 
Our method can be seen as preconditioned version of ADMM, and its convergence is accelerated by an improved update direction for each iteration.
O'Connor et al.~\cite{o2014primal} found ADMM outperforming PD for few iteration counts while using a restricted version of PD which does not exploit PD's full potential of optimal update directions.
We demonstrate the advantages of our approach over both ADMM and IOP in \myrefsec{sec:guidingResults}.

\paragraph*{Guiding}
A limitation of fluid simulation for computer graphics is the issue of control. 
Simply changing the resolution of a simulation can alter its behavior significantly.
A popular class of methods circumvents this problem by synthesizing smaller
scales in a decoupled fashion \cite{bridson2007curl,kim2008wavelet,pfaff2009synthetic}.
As a consequence, the details are easy to fine-tune, but do not tightly integrate with the base flow,
which is left unmodified. 

Fluid guiding is a challenging example of an inverse problem for fluids.
Adjoint methods have been used in engineering and graphics~\cite{giles2000introduction,McNamaraAdjointMethod} but require differentiation of the entire fluid solver.  
Guide shapes~\cite{nielsen2011guide} are able to add detail to free surface flows by applying velocity conditions to high-resolution simulations distant to the interface but are limited in volumetric contexts.
Other approaches aiming for high-level control have proposed the use of 
Lagrangian coherent structures \cite{Yuan2011} or sketches \cite{PanHTZB13} to give users intuitive controls.
As our guiding technique takes an arbitrary flow field as input to calculate
plausible and tightly coupled detail, such approaches would be a good complement for our method.

While Gregson et al.~\cite{GregsonCapture} also demonstrate preliminary results for guided flows, their approach
employed the fast Fourier transform for filtering low-frequencies and pressure-projection. This is a very efficient approach, but it becomes
impractical for more complex geometries and spatially varying guiding.

A simpler approach to guiding is via control
forces~\cite{ShiTamingLiquids,fattal2004target} or
velocities~\cite{shi2005controllable}. These approaches can result in
artificially viscous flows, as noted by Thuerey et al.~\cite{Thuerey:2006:SCA06dpfc}, who
proposed a multi-scale approach based on control particles that controls
low-spatial frequencies.  Possibly the most similar approach to ours is the one
by Nielsen et al.~\cite{NielsenGuidingSmoke,nielsen2010improved} who use a
multi-scale formulation based on low pass filters. They arrive at a monolithic
system with four degrees of freedom per cell, requiring a specialized solver to
reduce run time. In contrast, our control scheme decouples into separate and
more easily solved subproblems via recent developments in non-smooth
optimization.
Furthermore, for practical purposes, we introduce an \emph{upres} method into our workflow~\cite{kim2008wavelet,Yuan2011,previewBasedSampling2011,flowFieldMod2013,Huang2013} to facilitate the separation of low-resolution and high-resolution guiding.

\paragraph*{Boundary Conditions}
play a crucial role in fluid simulation, and as such have received a significant amount of attention. 
Foster et al. \cite{foster2001practical} describe how to allow for tangential motions of liquids,
and are the first to note problems with liquid unnaturally sticking to domain boundaries.
Batty et al. \cite{Batty2007} propose to solve inequality constraints with Quadratic Programming,
while Chentanez et al. \cite{ChentMG} observe that they can incorporate these inequalities into their multigrid solver.
Methods such as the FFT-based one of R. Henderson~\cite{Henderson2012} are similar in spirit
to our method, as they separate the BCs from the divergence-free projection step,
however, without targeting separating boundaries.
Other approaches realize unilateral incompressibility to allow for separation effects \cite{narain2010},\cite{narain2009}, \cite{gerszewski2013physics} but they also require complex
solvers to solve their proposed Quadratic Programming Problems. We will demonstrate how to solve for 
separating motions with a regular conjugate gradient (CG) solver 
based on our modular optimization framework.

\section{Methodology}
\label{sec:methodology}

\newcommand{\Pre}{p}

In graphics, fluids are typically simulated by solving the incompressible Euler equations,
written as
\begin{eqnarray}
\frac{\partial \u}{\partial t} + \u \cdot \nabla \u &=& - \nabla \Pre + \vec{f}_{\text{ext}} \ ,  \label{eq:NSE} \\
\nabla \cdot \u &=& 0, \label{eq:IncompCons} 
\end{eqnarray}
where $\u$ is the flow velocity, $\Pre$
the pressure, and $\vec{f}_{\text{ext}}$ the external body forces. 
Simulations commonly proceed via operator splitting to satisfy both constraints.
First, using all but the pressure term in \myrefeq{eq:NSE}, an intermediate velocity field is computed.
Then a pressure projection is applied to satisfy the divergence-free condition in \myrefeq{eq:IncompCons}. 
A pressure field that exactly counteracts the divergence is computed to correct the velocity field.
Orthogonality of the curl-free and divergence-free components ensures that divergence-free components of the flow 
are not affected and allows the pressure projection to be interpreted as an Euclidean projection onto the space of divergence-free velocity fields.

The splitting approach closely resembles convex optimization approaches originally developed for imaging inverse problems and machine learning involving non-smooth or constrained objective functions.
We show how several of these approaches can be adapted to solve difficult problems in fluids.

\paragraph*{Convex Optimization}
aims to solve problems of the form
\begin{equation}
\begin{aligned}
& \underset{\x}{\text{minimize}}
& & h(\x),
\end{aligned}
\label{eq:min_hx}
\end{equation}
where $h$ is a convex function that may be non-smooth or even discontinuous.
If $h$ is the sum of two simpler functions (say $h = f + g$), then we have
\begin{equation}
\begin{aligned}
& \underset{\x}{\text{minimize}}
& & f(\x) + g(\x).
\end{aligned}
\label{eq:min_fx_gx}
\end{equation}
A number of recently developed algorithms target this type of problem by employing an iterative divide-and-conquer approach.
These algorithms are known as \emph{proximal methods} and are defined in terms of so-called \emph{proximal operators} (we use $\gv$ exclusively to denote the generic argument variable):
\begin{equation}
\begin{aligned}
\prox_{f,\sigma}(\gv) &:= \argmin_{\x}\left(f(\x) + \frac{\sigma}{2}\norm{\x-\gv}^2\right).
\end{aligned}
\label{eq:proxFsigma}
\end{equation}
One such algorithm is the PD method~\cite{pock2009algorithm}, which solves a slightly more general problem:
\begin{equation}
\begin{aligned}
& \underset{\x}{\text{minimize}}
& & f(K\x) + g(\x)
\end{aligned}
\label{eq:PD_type_problem}
\end{equation}
for some linear operator $K$.
PD solves the problem iteratively by providing a series of variable updates that terminate when $\z$ converges to the solution. 
The combination of $\x, \z$ and $\y$ ensures that $\z$ converges to the optimal value of Eq. \ref{eq:PD_type_problem}. 
The updates are given by
\begin{align}
\x^{k+1} &:= \prox_{f^*, 1/\sigma}(\x^k + \sigma K \y^k)\label{eq:PD_x_update_orig} \\
\z^{k+1} &:= \prox_{g, 1/\tau}(\z^k - \tau K^* \x^{k+1}) \\
\y^{k+1} &:= \z^{k+1} + \theta(\z^{k+1}-\z^k),
\end{align}
where $\{\sigma,\tau,\theta\}$ are parameters that affect convergence, $f^*$ and $K^*$ are the convex conjugates of $f$ and $K$, respectively.
For our problem, $K$ is simply the identity, which leads to $K^* = K^T = I$. Note that if additionally $\sigma,\tau,\theta=1$, the iterative update scheme reduces to ADMM. 
A more appropriate choice ($\sigma,\tau,\theta \neq 1$) leads to optimal control over convergence.
As for $f^*$, it is not necessary to compute it directly. The proximal operator can be transformed using Moreau's identity:
\begin{align}
\prox_{f^*, 1/\sigma}(\gv) &= \gv - \sigma \, \prox_{f, \sigma}(\gv/\sigma).
\end{align}
The variable updates are thus reduced to
\begin{align}
\x^{k+1} &:= \x^k + \sigma \y^k - \sigma \, \prox_{f, \sigma}(\tfrac{1}{\sigma}\x^k + \y^k)\label{eq:PD_x_update} \\
\z^{k+1} &:= \prox_{g, 1/\tau}(\z^k - \tau \x^{k+1}) \\
\y^{k+1} &:= \z^{k+1} + \theta(\z^{k+1}-\z^k).
\end{align}
A more in-depth discussion of PD can be found in~\cite{ChambollePD}.

The advantage of using proximal methods is that the optimization can be performed separately for the two objective functions, allowing difficult optimizations to be split into more manageable components.
Also, depending on the form of $f$ and $g$, many special cases can be significantly simplified~\cite{Boyd:ADMM} by exploiting their mathematical structures.

To the best of our knowledge, PD has not yet been applied to fluid problems, 
despite its provably optimal convergence properties for the class of problems of \myrefeq{eq:PD_type_problem}.
A pseudocode implementation of our PD-based optimization method is given in \myrefalg{alg:PDgeneric} for one time step. 
Comparing to previous methods, the improved
convergence is achieved with only a minimal increase in computational cost. 
A few more vector additions
and multiplications are necessary, which typically are negligible compared to the cost of
the proximal operators.
We demonstrate the convergence of our method and compare it to other methods in \myrefsecs{sec:guidingResults}{sec:resultsBC}.
For reference, we briefly review IOP and ADMM in \myrefapp{app:ADMMIOP}.

\paragraph*{Convex Optimization of Fluids}
In fluid simulation, we often encounter problems of the form
\begin{equation}
\begin{aligned}
& \underset{\x}{\text{minimize}}
& & f(\x) \\
& \text{subject to}
& & \x \in \cdiv,
\end{aligned}
\label{eq:fluids_problem}
\end{equation}
where $\x$ is the velocity field we seek, and $\cdiv$ is the space of divergence-free velocity fields.
$f$ must be a convex function. Practically, this can be either a quantity we are trying to minimize, or a hard constraint that must be satisfied (in which case, $f$ would be an indicator function).

The second constraint requires $\x$ to be divergence-free. 
Removing the divergent part of the flow can also be viewed as an orthogonal projection~\cite{chorin1968numerical,parikh2013proximal}.
Gregson et al.~\cite{GregsonCapture} made the key observation that the proximal operator 
for $\cdiv$ can be easily computed via a pressure projection.  In other words, we have
\begin{equation}
\prox_{g,1/\tau}(\gv) = \PiDiv(\gv),
\label{eq:proxG}
\end{equation}
where $\PiDiv$ denotes a projection onto $\cdiv$ with a commonly used Poisson solver.
Hence, formulating a fluid problem this way allows the optimization algorithm to be easily integrated into a common fluid solver---we simply replace the call for a pressure projection subroutine 
with a call to the PD optimization step outlined in \myrefalg{alg:PDgeneric}. 
We check for convergence using a threshold parameter $\epsilon$, and stop the algorithm once the per-iteration change of $\z$ falls below this threshold (\myrefalg{alg:PDgeneric}, line 12).
Note that we define the pressure projection to include only the calculation of the pressure values,
and a subtraction of the pressure gradient from $\x$, excluding any optional modifications of the velocity field.

\begin{algorithm}[b!]
	\caption{Our PD-based method for fluid simulation}
	\label{alg:PDgeneric}
	\begin{algorithmic}[1]
		\Procedure{PD}{$\uc$, $\ut$, $\tau$, $\sigma$, $\theta$}
		\While {$k < maxIters$}
		\State // $\x$-update
		\State $\x^{k+1}
		\leftarrow \x^k + \sigma \y^k - \sigma \mathrm{proxF}(\sigma,\tfrac{1}{\sigma}\x^k+\y^k)$
		\State // $\z$-update (using adaptive CG accuracy)
		\State $\z^{k+1} \leftarrow \PiDiv(\z^k - \tau \x^{k+1})$
		\State // $\y$-update
		\State $\y^{k+1} \leftarrow \z^{k+1} + \theta (\z^{k+1} - \z^k)$
		\State // check stopping criterion
		\State $\r^{k+1} \leftarrow \z^{k+1} - \z^k$
		\State $\epsilon \leftarrow \sqrt{\nDim} \epsAbs + \epsRel \norm{\z^{k+1}}$
		\State \textbf{if}\, {$\left(\norm{\r^{k+1}} \leq \epsilon\right)$} \textbf{then break}
		\EndWhile
		\State \Return $\z$
		\EndProcedure
	\end{algorithmic}
\end{algorithm}

The pressure projection implemented by a CG solver is usually the most expensive part in regular fluid simulation, and its effect is magnified in our algorithm due to its iterative nature.
However, this iterative set-up gives us an opportunity to apply an adaptive scheme.
Let $\epsCG$ be the accuracy of the CG solver.
We choose its value starting with a high threshold (e.g., $\epsCG=10^{-2}$) and then adaptively decrease it over the PD iterations. 
We decrease $\epsCG$ as soon as the per-iteration change of $\z$ is close to $\epsilon$; until $\epsCG$ reaches the desired final accuracy (e.g., $\epsCG=10^{-5}$).
Using this adaptive CG scheme greatly accelerates the performance by cutting down on CG iterations in the beginning of the optimization when the divergence-free constraint does not need to be strictly enforced.

\myrefalg{alg:PDgeneric} summarizes the general framework of our method as it applies to fluid simulation.
Specific applications call for different definitions of $f$, which in turn affects how the $\x$-update is computed.
In the following sections, we discuss how to apply this method to the
fluid guiding and the separating BC problem.
For each application, we define the appropriate $f$ and discuss how to compute its proximal operator.

\section{Fluid Guiding}
\label{sec:fluid_guiding}

In fluid guiding, the goal is to minimize the change applied to the current velocity field such that the resulting velocity field follows the large-scale motions of a given target velocity. 
The objective function $f$ is given by
\begin{align}
f(\x) &= \norm{\blurM(\x-\ut)}^2 + \norm{\SVW(\x-\uc)}^2,
\label{eq:main_objective}
\end{align}
where $\uc$ is the current velocity field (after advection and before pressure projection), $\ut$ is the target velocity field, $\x$ is the guided velocity field, 
$\SVW$ is the guiding weights matrix, and $\blurM$ is the Gaussian blur matrix. 
The first term of this objective function is minimal for a solution that matches the target velocity when blurred by $\blurM$, while the second term penalizes solutions far away from the current flow field.

In order to keep the application general, both matrices in our objective function are spatially varying (Nielsen and Christensen~\cite{nielsen2010improved} also performed fluid guiding with spatially varying guiding weights, 
but not with spatially varying blur). 
$\SVW=\SVW(\x)$ is a diagonal matrix containing spatially varying weights that control the guiding strength (larger entries denote weaker guiding), 
and $\blurM=\blurM(\blurR,\x)$ is the Gaussian blur matrix that applies a blur of radius $\blurR$ only to fluid cells so that we can handle boundaries and obstacles in our domain.

Our objective function in \myrefeq{eq:main_objective} is quadratic. That is, it can be expressed as
\begin{align}
f(\x) &= \half \x^T A \x + \vec{b}^T \x + c,
\end{align}
where
\begin{align}
A &= 2(\blurM^T \blurM + \SVW^2) \\
\vec{b} &= -2(\blurM^T \blurM \ut + \SVW^2 \uc)\\
c &= \ut^T \blurM^T \blurM \ut + \uc^T \SVW^2 \uc.
\end{align}
Note that $\SVW^T \SVW = \SVW^2$ since $W$ is symmetric.

We can combine $f'(\x)=\vec{0}$ (three equations per cell in 3D) and the divergence-free constraint (one equation per cell) into a linear system $L \x = \vec{d}$. 
But since this system is overconstrained, we solve it in the least-squares sense by considering
\begin{align}
L^T L \x&= L^T \vec{d},
\end{align}
\label{eq:directSolve}
where $L = \left( \begin{smallmatrix} A \\ \nabla\cdot \end{smallmatrix} \right)$ and $
\vec{d} = \left( \begin{smallmatrix} \vec{b} \\ \vec{0} \end{smallmatrix} \right)$.
This quadratic system can be solved using an iterative solver such as the CG solver. 
However, this approach is infeasible since the number of elements in the matrix grows by $\mathcal{O}(N)$, where $N$ is the volume of the simulation domain. 
Thus PD is still a good choice even for seemingly simple quadratic energies. We will see later how this direct CG solver compares to our algorithm in terms of performance.

Instead, we make use of the quadratic property of $f$ to simplify its proximal operator to
\begin{align}
\prox_{f,\sigma}(\gv) &= (A + \sigma I)^{-1}(\sigma \gv - \vec{b}).
\label{eq:quad_form_proxF}
\end{align}
Factoring out $\uc$, we obtain
\begin{align}
\prox_{f,\sigma}(\gv) = \uc + M^{-1}(\sigma \gv  +\q),
\end{align}
where
\begin{align}
M &= 2 \blurM^T \blurM + 2 W^2 + \sigma I\\
\q &= 2 \blurM^T \blurM(\ut - \uc) - \sigma \uc.
\end{align}

\begin{figure*}[thb!]
	\centering
	\begin{subfigure}{0.17\textwidth}
		\includegraphics[width=\textwidth]{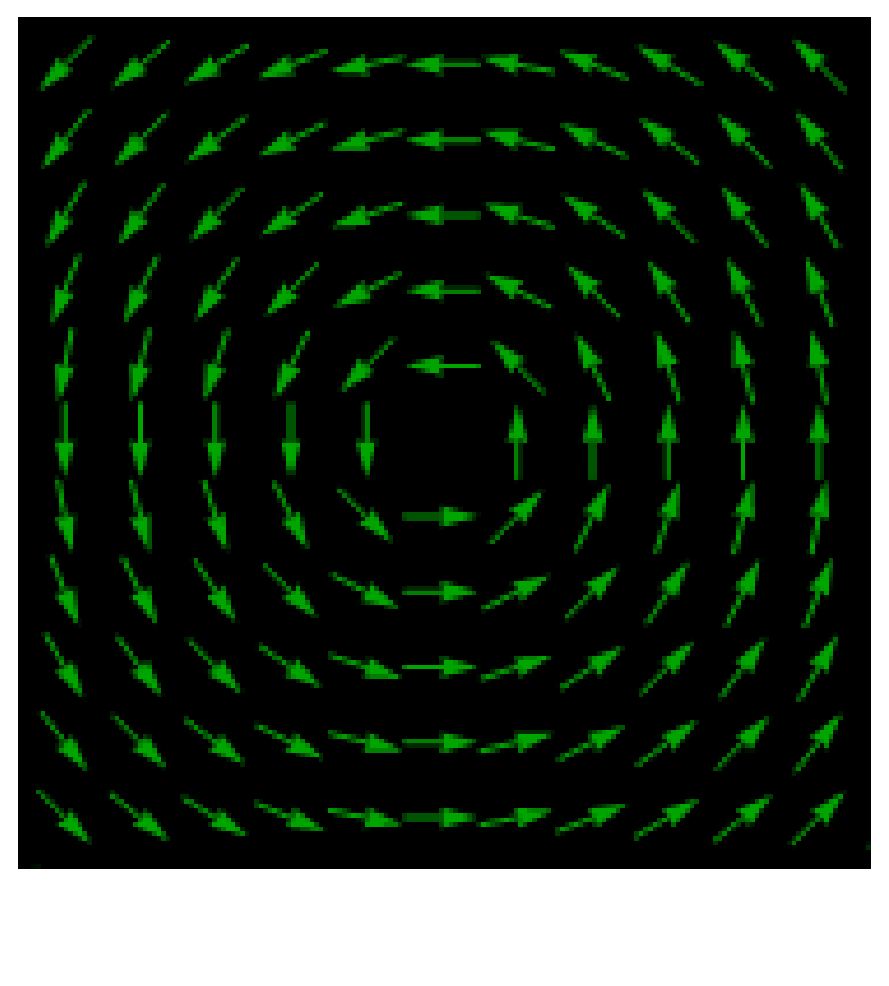}
		\caption{Target velocity}
		\label{fig:circular_target}
	\end{subfigure} \quad
	\begin{subfigure}{0.34\textwidth}
		\includegraphics[width=\textwidth]{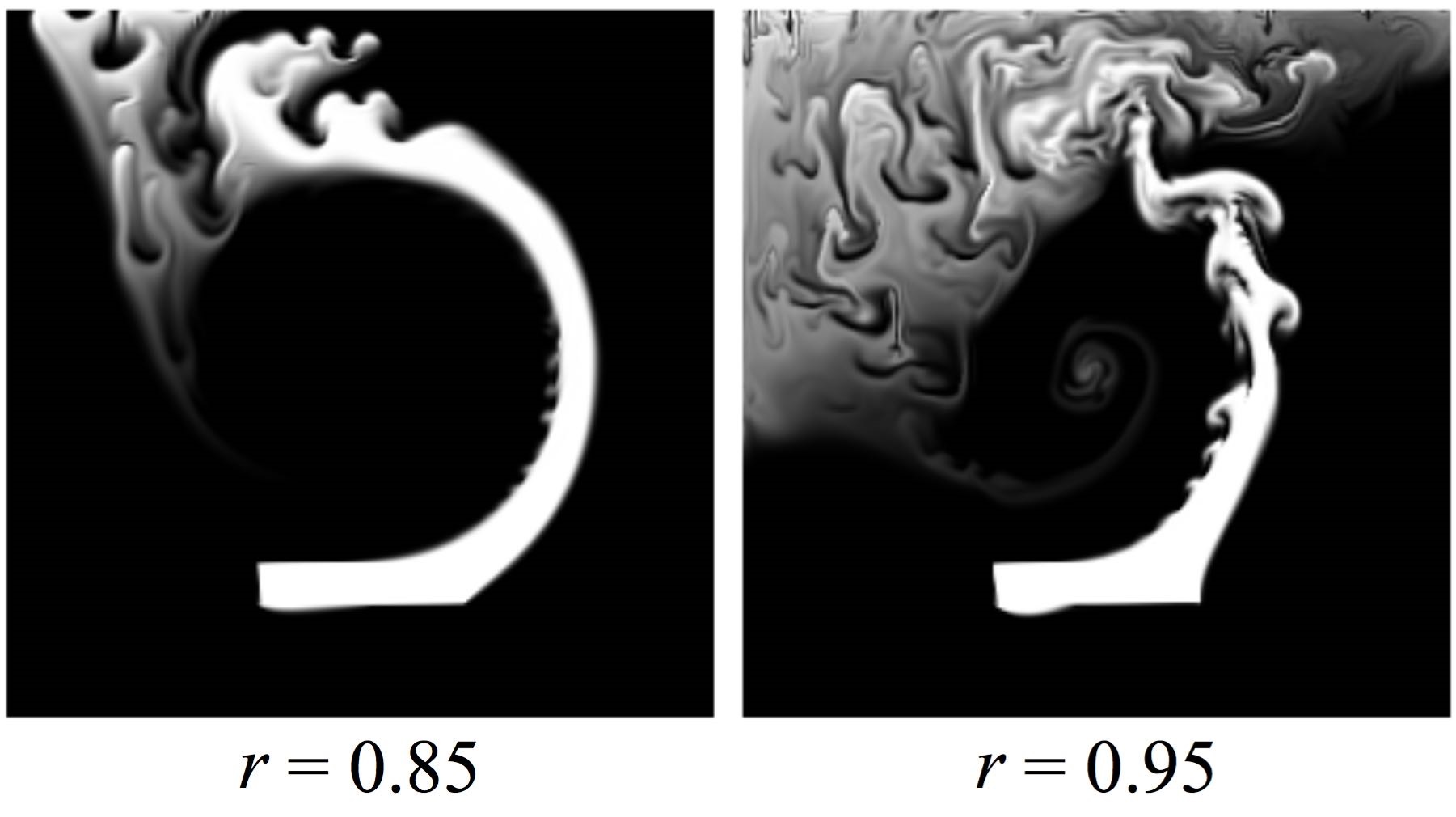}
		\caption{Linear velocity blend}
		\label{fig:linear_blend}
	\end{subfigure} \quad
	\begin{subfigure}{0.34\textwidth}
		\includegraphics[width=\textwidth]{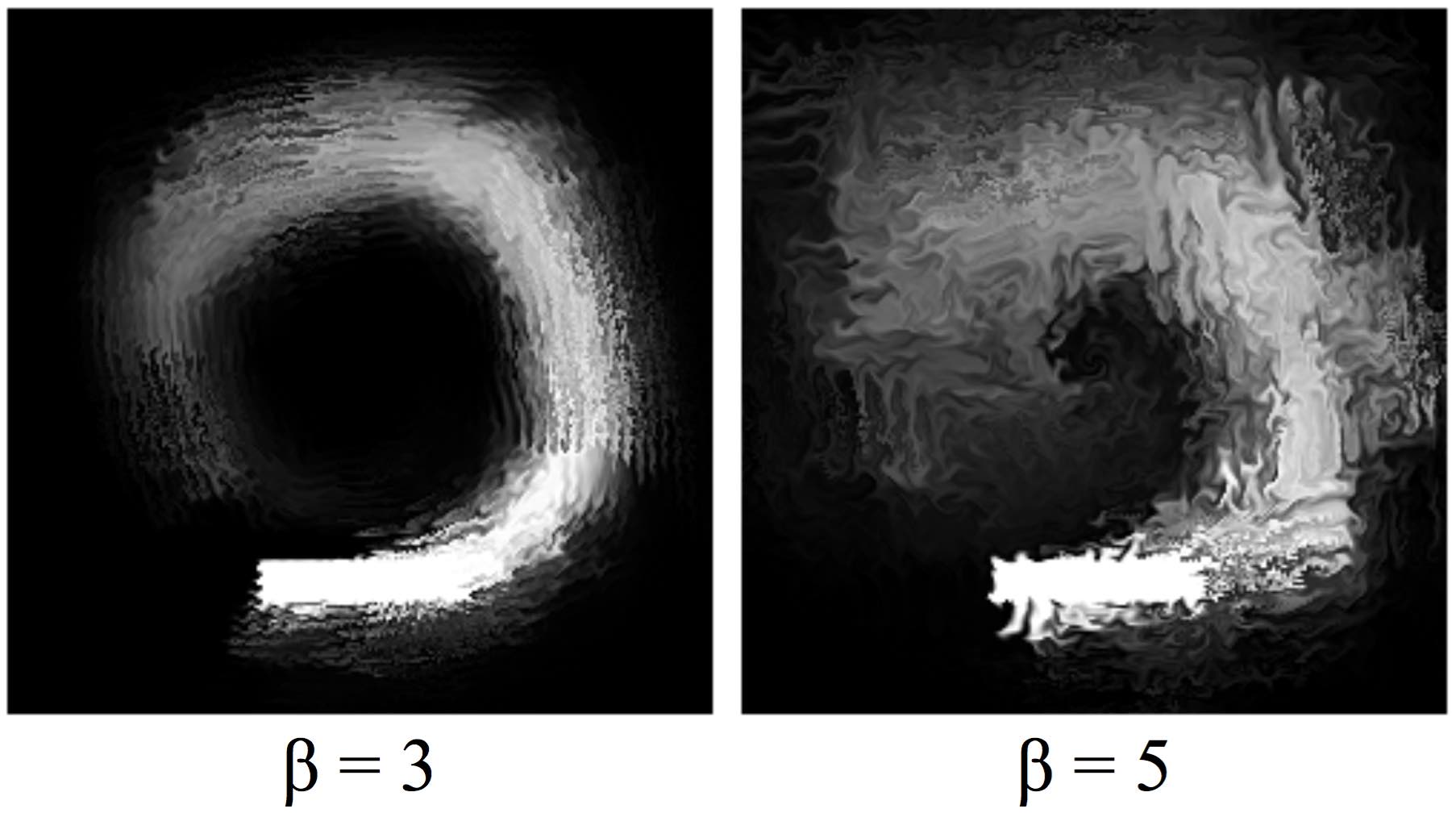}
		\caption{Detail-preserving blend}
		\label{fig:detail_blend}
	\end{subfigure}
	\vspace{-1mm}
	\caption{\Naive guided simulations with circular targets.}
	\label{fig:naive_blend}
\end{figure*}

Computing $M^{-1}(\sigma \gv  +\q)$ for every iteration is slow. 
Instead, we precompute $\q$ and $M^{-1}$ since they do not rely on previous iterations. 
In fact, using the Sherman-Morrison-Woodbury formula, $M^{-1}$ can be approximated as
\begin{align}
\label{eq:invM_approx}
M^{-1} &\approx (2 \SVW^2 + \sigma I)^{-1} - 2 (2 \SVW^2 + \sigma I)^{-1} \blurM^T \blurM (2 \SVW^2 + \sigma I)^{-1}.
\end{align}
See \myrefapp{app:invM_derivation} for derivation details.

Lastly, spatially invariant Gaussian blurs are symmetric, so $\blurM^T \blurM = \blurM^2$. 
They are also separable, and can be applied independently in each dimension. 
This combined with the matrix inversion approximation greatly speeds up the $\x$-update.
For Gaussian blurs with spatial variation, we still use a symmetric and hence separable Gaussian blur (with different blur radii) for each cell, but $\blurM$ itself is no longer symmetric.
However, if the spatial variation is limited (e.g. different blur radii on the left and right sides of the simulation domain), 
we approximate $\blurM^2$ by $\blurM^T \blurM$, since $\blurM^T \blurM \approx \blurM^2$ for regions of constant blur. 
We found our approximation to work well in practice, and we will show examples using this approximation later.
Our PD guiding scheme is independent of the particular choice and implementation of the blur kernel. 
Given a fast implementation for calculating $\blurM^T$, it would be straight-forward to extend our method with a full evaluation of $\blurM^T \blurM$.

Additionally, our approach for applying the blur kernel easily allows for interior boundaries by using a blur radius of zero for obstacle cells. 
This is a key difference from guiding scheme based on the fast Fourier transform~\cite{GregsonCapture}, which typically require periodic domains without internal boundaries. 

\myrefalg{alg:proxF} summarizes how $\prox_{f,\sigma}(\gv)$ is approximated. 
Note that the two extra parameters $\SVW$ and $\blurR$ in \myrefalg{alg:proxF} as compared to \myrefalg{alg:PDgeneric} are specific to the guiding application.

\begin{algorithm}
	\caption{Approximation for $\prox_{f,\sigma}(\gv)$ in fluid guiding}
	\label{alg:proxF}
	\begin{algorithmic}[1]
		\Procedure{proxF}{$\sigma$, $\gv$, $\SVW$, $\blurR$}
		\State $\q = 2 \blurM^T \blurM (\ut - \uc) - \sigma \uc$ // can be precomputed
		\State $\gamma = (2 \SVW^2+\sigma I)^{-1}$ // inverse of diagonal matrix
		\State \Return\, $\uc + \gamma(\sigma \gv + \q) - 2 \blurM^T \blurM \gamma^2 (\sigma \gv + \q)$
		\EndProcedure
	\end{algorithmic}
\end{algorithm}

\paragraph*{Evaluation}
\label{sec:FG_eval}
In this section, we will first demonstrate the efficacy of our method using 2D examples, followed by more impressive 3D results. 
To simplify the notation, when the guiding weight is spatially invariant (say, $c$ everywhere), we will write $\SVW=c$. 
And when the guiding weight is spatially varying (say, $c_1$ and $c_2$ on the left and right side on the simulation domain, respectively), we will write $\SVW=(c_1,c_2)$. 
Similarly, spatially varying blur radii would be written as $\blurR = (r_1,r_2)$.

We first compare our method to two \naive guiding methods, with a counterclockwise circular velocity field as the target $\ut$, as shown in \myreffig{fig:circular_target}. 
Linear velocity
blend computes the new velocity as a linear combination of the current velocity and the target velocity: \mbox{$
	\u_{\new} = \ratio \uc + (1-\ratio) \ut$},
where \mbox{$0 \leq r \leq 1$}.
Detail-preserving guiding \cite{Thuerey:2006:SCA06dpfc} subtracts the large-scale (i.e., blurred) motions from the current velocity before adding the target velocity: \mbox{$\u_{\new} = \uc - \blurM \uc + \ut$}.
The idea is to create a new velocity with large-scale motions from $\ut$ and small-scale details from $\uc$.
Both methods are followed by a pressure solve to ensure zero divergence.

With linear velocity blend, guiding strength is increased with smaller $\ratio$. 
The method's main weakness is that small fluid details are smoothed out with strong guiding, and more details come at the expense of reduced trajectory control (see \myreffig{fig:linear_blend}).

Detail-preserving blend allows guiding of large-scale motions towards the target velocity while preserving details. 
However, the details are difficult to control and have an unnatural frosted glass appearance (see \myreffig{fig:detail_blend}).

\begin{figure}[tb!]
	\centering
	\includegraphics[width=0.48\textwidth]{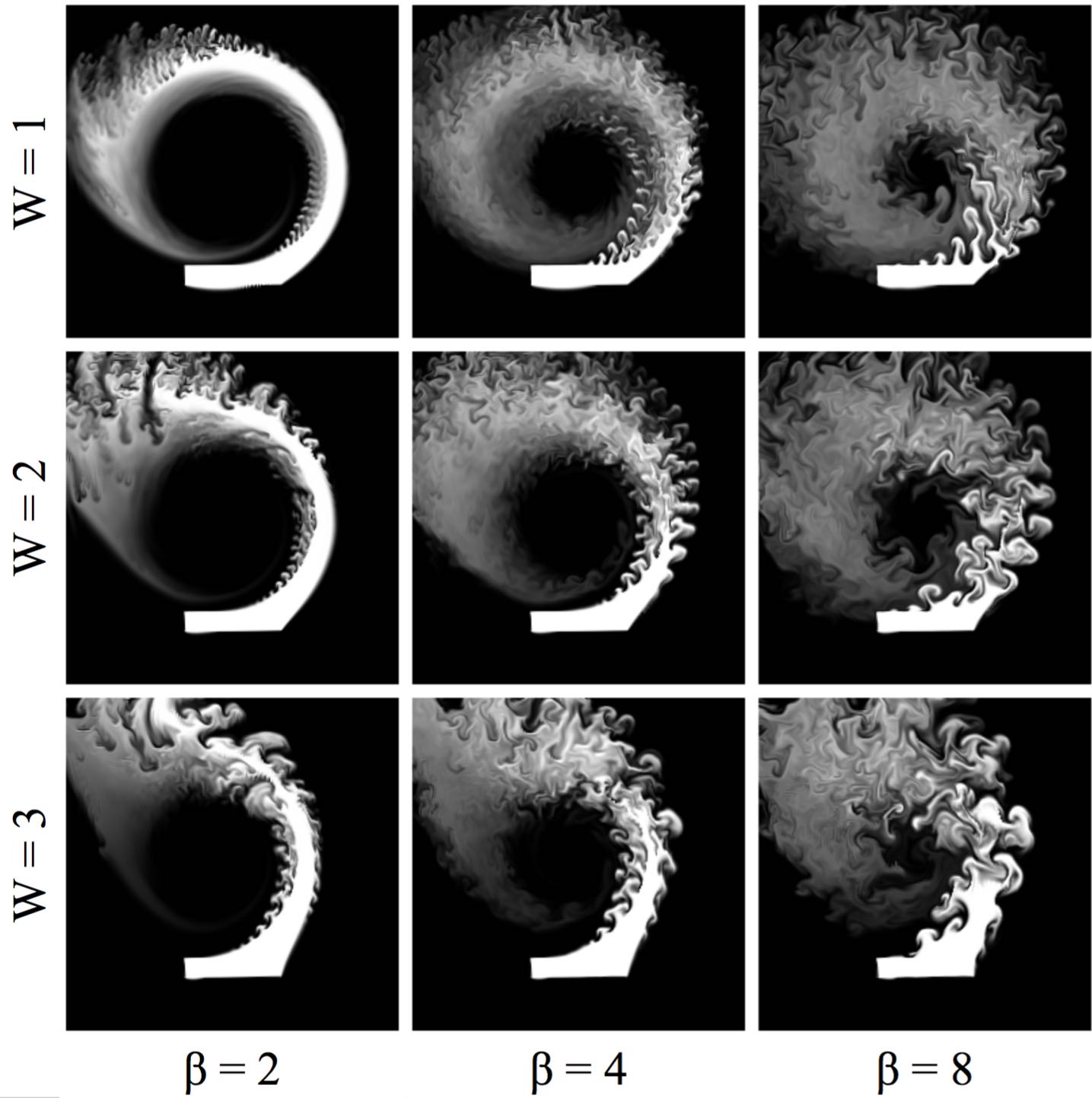}
	\caption{Guided simulations with circular targets, using our method with varying $\SVW$ and $\blurR$ for multi-scale motion control.
	}
	\label{fig:alphaVar_blurVar}
\end{figure}

In contrast, our method has much more flexible motion control when applied to fluid guiding, with $\SVW$ affecting the large-scale guiding strength and the blur radius $\blurR$ controlling the small-scale details. 
\myreffig{fig:alphaVar_blurVar} shows how the parameters affect guiding for a 2D smoke simulation with a circular target velocity.
Notice that larger $\SVW$ values allow for more freedom to deviate from the target, while larger $\blurR$ values lead to the formation of larger vortices.

\begin{figure}[tb!]
	\centering
	\includegraphics[width=0.42\textwidth]{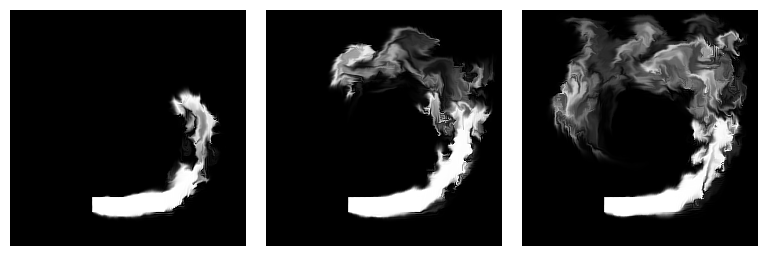}
	\includegraphics[width=0.42\textwidth]{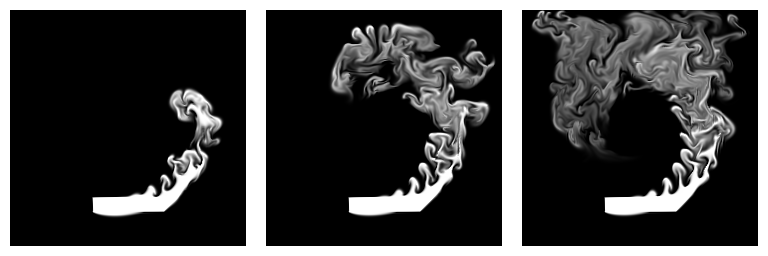}
	\caption{Upsampling from $64^2$ to $256^2$ using wavelet turbulence (top) and using our method (bottom). 
		Our method yields significantly more coherent vortices.
	}
	\label{fig:waveletTurbulence}
\end{figure}

We also compare our algorithm to wavelet turbulence \cite{kim2008wavelet}, as shown in
\myreffig{fig:waveletTurbulence}.
Although wavelet turbulence is fast---since it is purely a postprocessing technique---the resulting vortices do not couple as tightly and realistically as with our method.

\paragraph*{Performance}

\begin{figure}[tb!]
	\centering
	\includegraphics[width=0.37\textwidth]{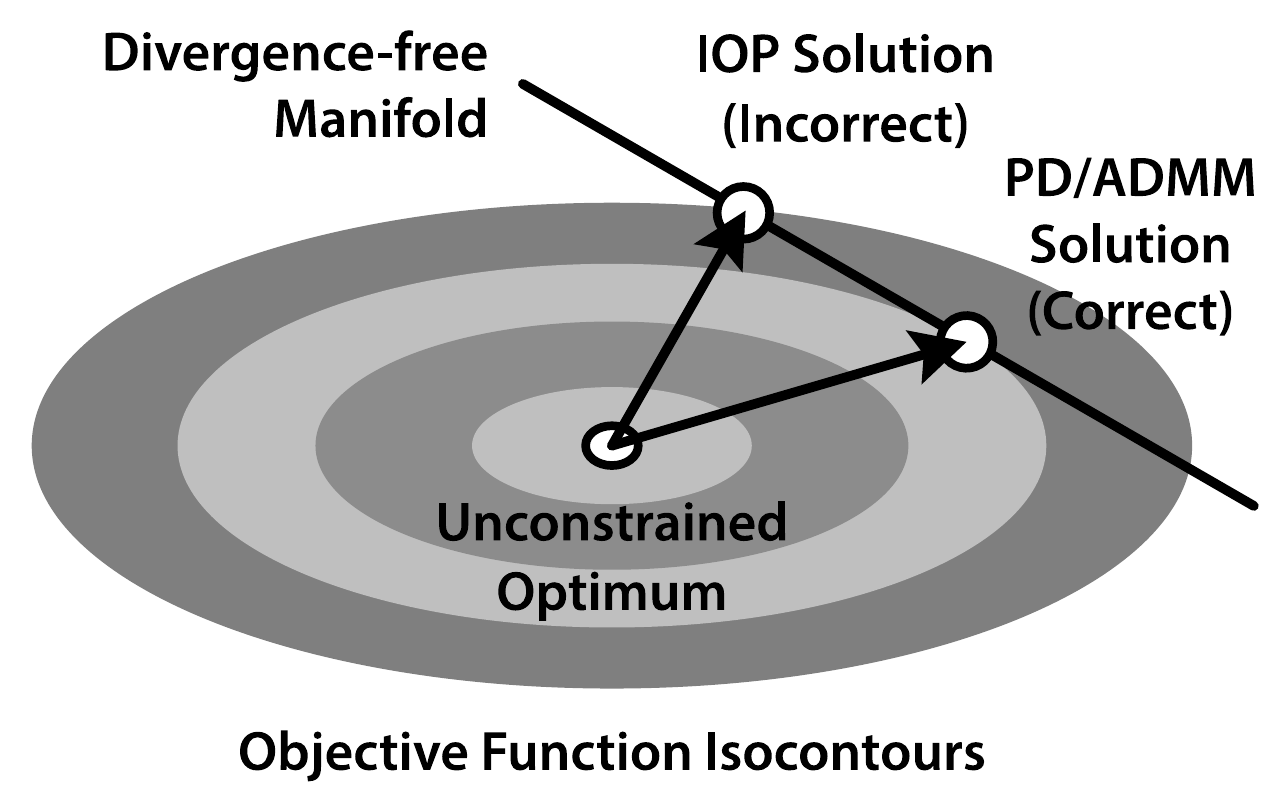}
	\caption{A divergent target velocity leads to IOP failure, while ADMM and PD converge successfully. 
		In this case, even though the IOP solution is divergence-free, it lands on an isocontour farther from the center than the PD/ADMM solution, which is the true minimum.}
	\label{fig:IOP_fail}
\end{figure}

Next, we examine our method in relation to other proximal methods, namely, ADMM and IOP (see \myrefapp{app:ADMMIOP} for details). 
ADMM is fairly straightforward to apply due to its similarity to PD. 
IOP can be interpreted as a fluid-specific version of the Projection onto Convex Sets (PoCS) algorithm~\cite{bauschke1996projection}, 
which solves convex feasibility problems (locating an intersection point of convex sets) rather than the more general problem solved by ADMM and PD. 
Applying IOP \naively alternates between the minimizer of $f$ and its closest divergence-free neighbor but does not yield the divergence-free minimizer of $f$, as illustrated by the failure case in \myreffig{fig:IOP_fail}.
In practice, we discovered instances for which IOP found plausible approximate solutions in a short time,
but its reliability is limited for general guiding applications. As such, we focus on only comparing our method to ADMM. 
All performance measurements were done on PCs with Intel Xeon E5-1650 CPUs.

For fairness of comparison, we experimented with several test scenes to deduce as-optimal-as-possible parameter sets for both ADMM and our method. 
These parameters, on average, produce the fastest convergence rates under various guiding weights and blur radii. 
The optimal parameters are correlated with the guiding weight; we analyzed their relationship and define the parameters in terms of the average guiding weight, 
$\bar{\SVW}$, which is simply the mean of the diagonal terms of $\SVW$.
For our method, we chose \mbox{$(\tau,\sigma,\theta)=(0.58/\bar{\SVW},2.44/\tau,0.3)$}, and for ADMM, \mbox{$\rho = 1.4\bar{\SVW}^2$}.
Before comparing their performance, it is important to note that we apply one of our contributions crucial for performance (the matrix inverse approximation and the separable Gaussians) to both methods. 

\begin{figure}[tb!]
	\centering
	\begin{subfigure}{0.45\textwidth}
		\includegraphics[width=\textwidth]{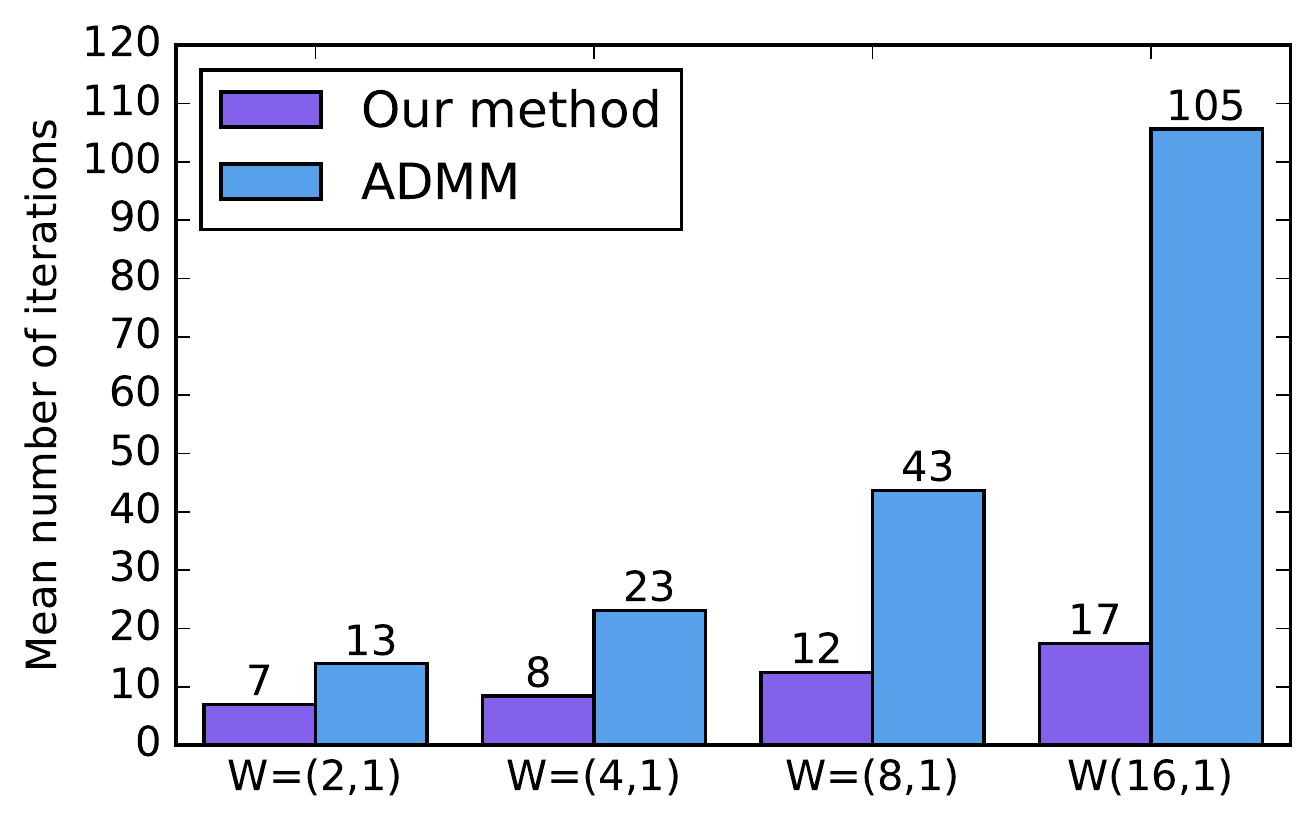}
		\caption{2D circular example}
		\label{fig:FG_timing_2D}
	\end{subfigure}\\
	\begin{subfigure}{0.45\textwidth}
		\includegraphics[width=\textwidth]{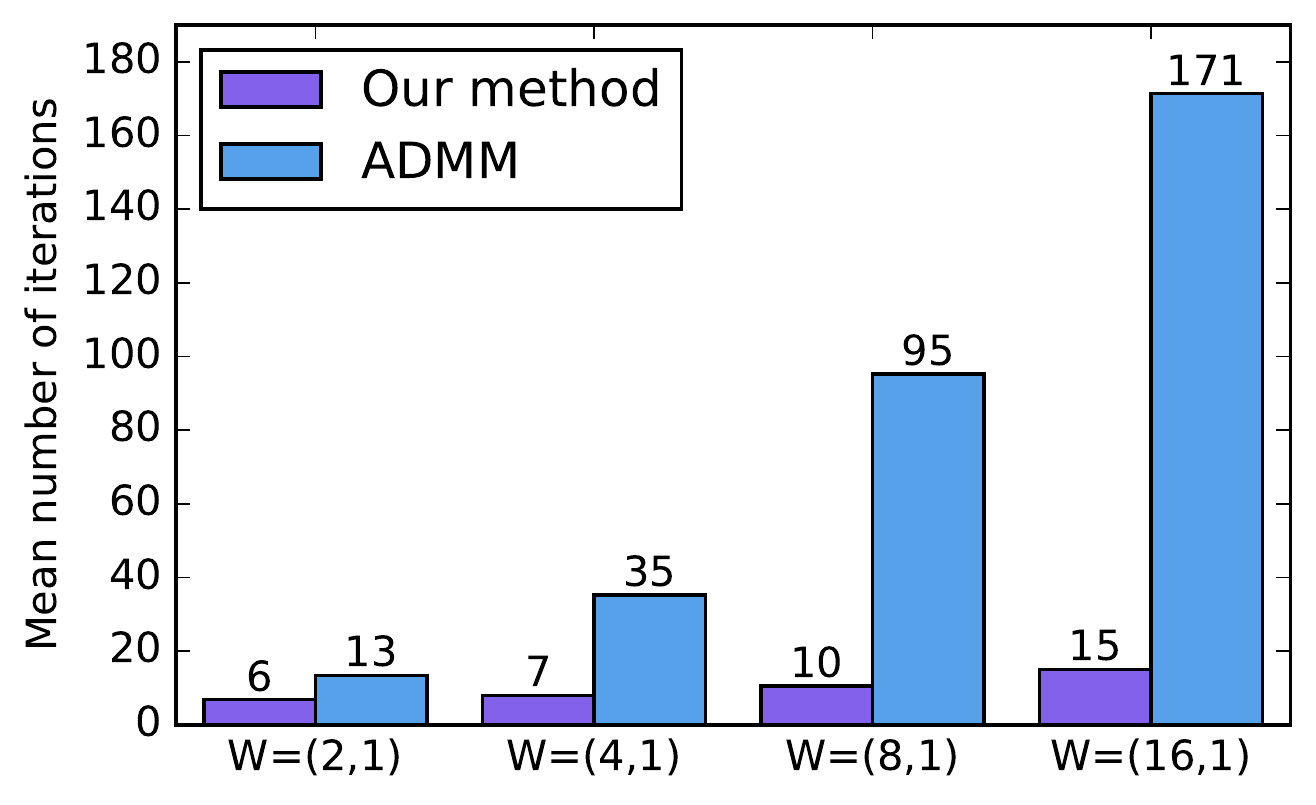}
		\caption{3D obstacle example}
		\label{fig:FG_timing_3D}
	\end{subfigure}
	\vspace{-1mm}
	\caption{Mean number of iterations per time step for guiding.}
	\label{fig:FG_timing}
\end{figure}

We compare the performance of ADMM and our method for 2D and 3D problems. 
The 2D problem is a $256^2$ simulation guided by a circular target velocity, similar to the one shown in \myreffig{fig:alphaVar_blurVar}, but with spatially varying weights and $\blurR=1$. 
Specifically, the guiding weight is fixed at 1 for the right side of the domain, while the left side takes values from \mbox{$\{2,4,8,16\}$}. 
\myreffig{fig:FG_timing_2D} compares, for the two methods, the mean number of iterations required to reach convergence. 
When the spatial variation is low, both methods perform decently. However, as the spatial variation increases, ADMM takes much longer to converge.

The 3D problem is a $120^3$ simulation with an obstacle, similar to the one in \myreffig{fig:obstacleSVW}, but with the same $\SVW$ and $\blurR$ parameters as the 2D problem. 
The performance results are shown in \myreffig{fig:FG_timing_3D}. Again, our method outperforms ADMM. In fact, the difference in convergence rates is even more apparent for the 3D example for large spatial variations. 
The run times for both ADMM and our method very closely correlate with the mean iteration counts. For brevity, the run times are given in \myreftab{tab:meanRuntime}.

\begin{table}[tb!]
	\centering
	\bgroup
	\def\arraystretch{1.2}
	\begin{tabular}{cc|c|c|c|c|l}
		\cline{3-6}
		& & \multicolumn{4}{ c| }{\textbf{Guiding weight \textit{W}}} \\ \cline{3-6}
		& & \textbf{(2,1)} & \textbf{(4,1)} & \textbf{(8,1)} & \textbf{(16,1)} \\ \cline{1-6}
		\multicolumn{1}{ |c  }{\multirow{2}{*}{\textbf{2D}} } &
		\multicolumn{1}{ |c| }{\textbf{Our method}} & 0.14 & 0.17 & 0.26 & 0.38     \\ \cline{2-6}
		\multicolumn{1}{ |c  }{}                        &
		\multicolumn{1}{ |c| }{\textbf{ADMM}} & 0.26 & 0.41 & 0.76 & 1.89     \\ \cline{1-6}
		\multicolumn{1}{ |c  }{\multirow{2}{*}{\textbf{3D}} } &
		\multicolumn{1}{ |c| }{\textbf{Our method}} & 16.6 & 19.5 & 26.3 & 38.4  \\ \cline{2-6}
		\multicolumn{1}{ |c  }{}                        &
		\multicolumn{1}{ |c| }{\textbf{ADMM}} & 30.4 & 67.8 & 176.9 & 325.6  \\ \cline{1-6}
	\end{tabular}
	\egroup
	\caption{Mean run time (in seconds) per time step for guiding.}
	\label{tab:meanRuntime}
\end{table}

\begin{figure}[tb!]
	\centering
	\begin{subfigure}{0.26\textwidth}
		\includegraphics[width=\textwidth]{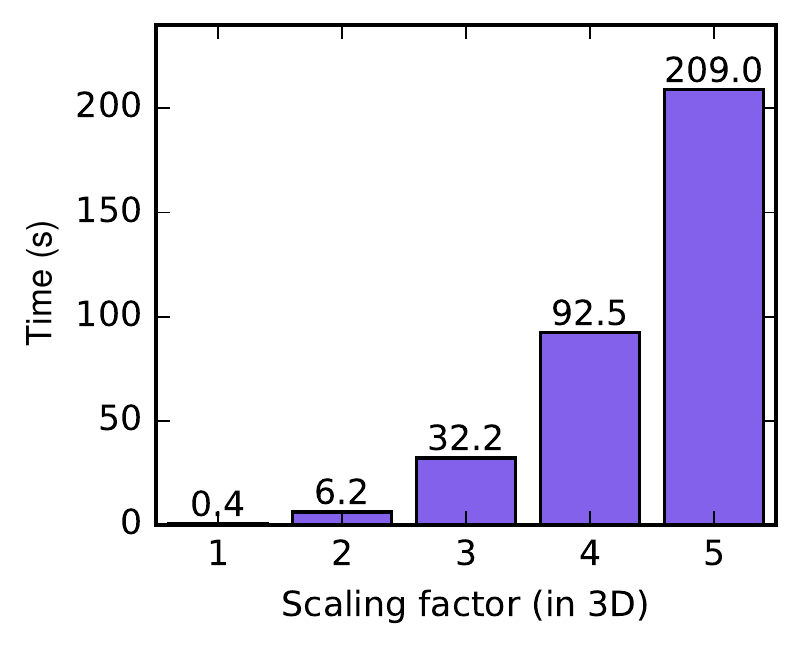}
		\caption{}
		\label{fig:res_times_total}
	\end{subfigure}
	\begin{subfigure}{0.21\textwidth}
		\includegraphics[width=\textwidth]{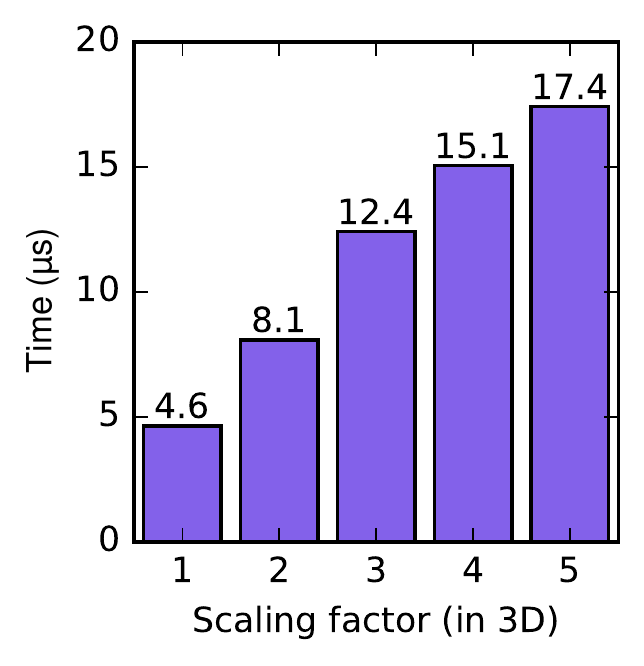}
		\caption{}
		\label{fig:res_times_per_cell}
	\end{subfigure}
	\caption{Performance of our method on the 3D tornado example at various resolutions, including (a) the mean run time per time step, and (b) the mean run time per time step per grid cell.}
	\label{fig:res_timing}
\end{figure}

In addition to the comparison with ADMM, we also analyze how our method scales as the resolution increases. 
In \myreffig{fig:res_times_total}, we run the guided 3D tornado simulation shown in \myreffig{fig:teaser} at five different resolutions and compare the mean run time per time step. 
The scaling factor is with respect to the lowest resolution $40^2 \times 60$; for instance, a scaling factor of 3 corresponds to the resolution $120^2 \times 180$. 
Notice that even at the largest resolution $200^2 \times 300$, each time step takes less than four minutes to complete. 
However, since the run times appear to scale exponentially with respect to resolution, we plot the mean run time per grid cell (see \myreffig{fig:res_times_per_cell}) to show that the scaling is in fact linear.

Previously, we mentioned that applying a CG solver also works for the guiding problem, although its poor performance renders it infeasible. 
To test this, we ran a small $128^2$ example with a circular target velocity field and found the direct CG solver (see \myrefeq{eq:directSolve}) approach to be 4000 times slower than our method.

The performance tests above are all done with our generic guiding method, which handles spatially varying weights and blurs. 
In the special case with spatially invariant weights, we can apply a non-iterative method to greatly speed up the simulation. 
The method involves defining $f(\x)$ in terms of the non-divergent components of the input velocities and then finding its minimizer. 
Since differentiation commutes with convolution, the result will be automatically divergence-free for spatially invariant weights. 
Although this non-iterative method offers potential speed-up, we focus on the general case of spatially varying operators that requires iterating.

\subsection{Guiding Results}
\label{sec:guidingResults}

\begin{figure}[tb!]
	\centering
	\begin{subfigure}{0.23\textwidth}
		\includegraphics[width=\textwidth]{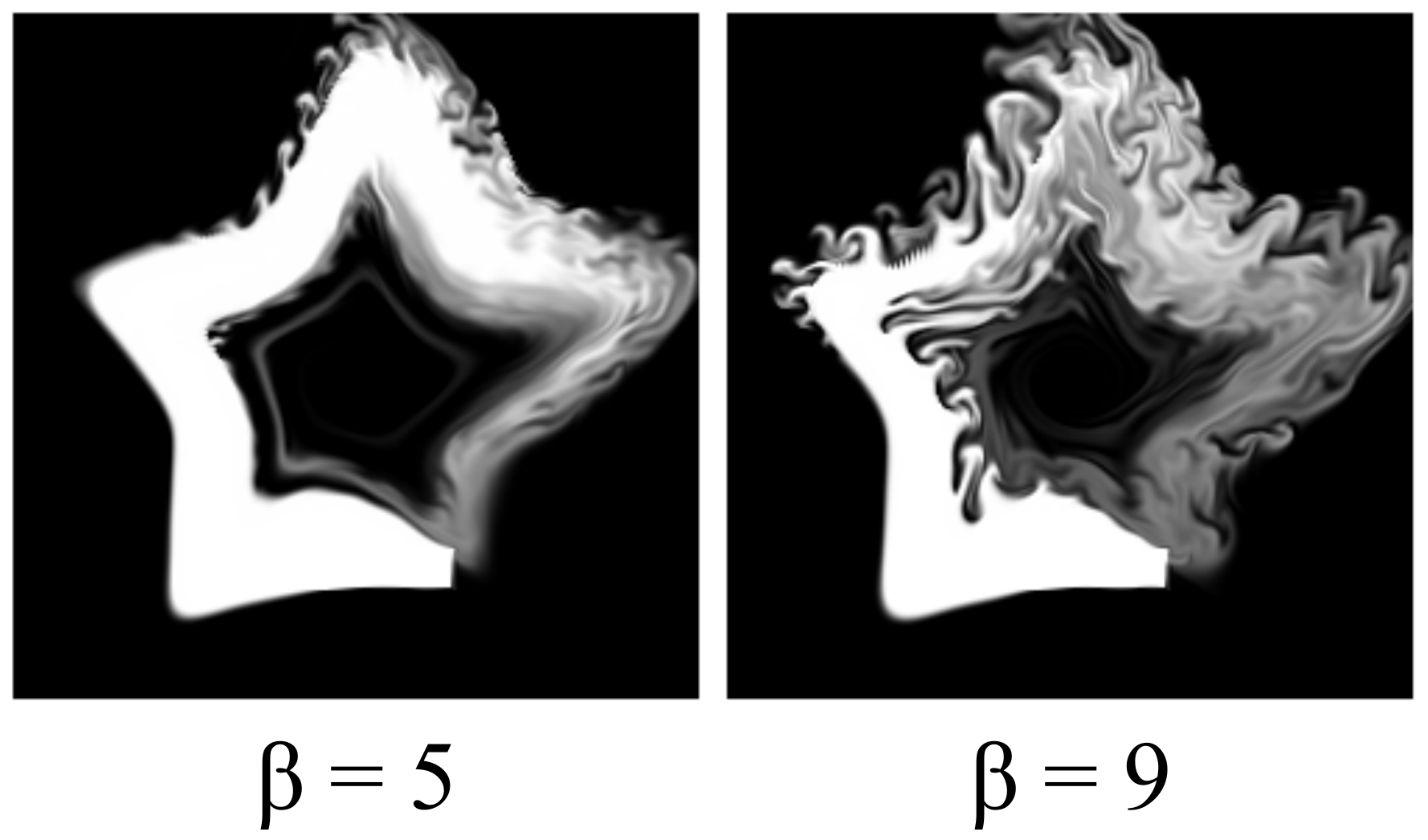}
		\caption{with up-res}
		\label{fig:starGuided_upsample}
	\end{subfigure}\,
	\begin{subfigure}{0.23\textwidth}
		\includegraphics[width=\textwidth]{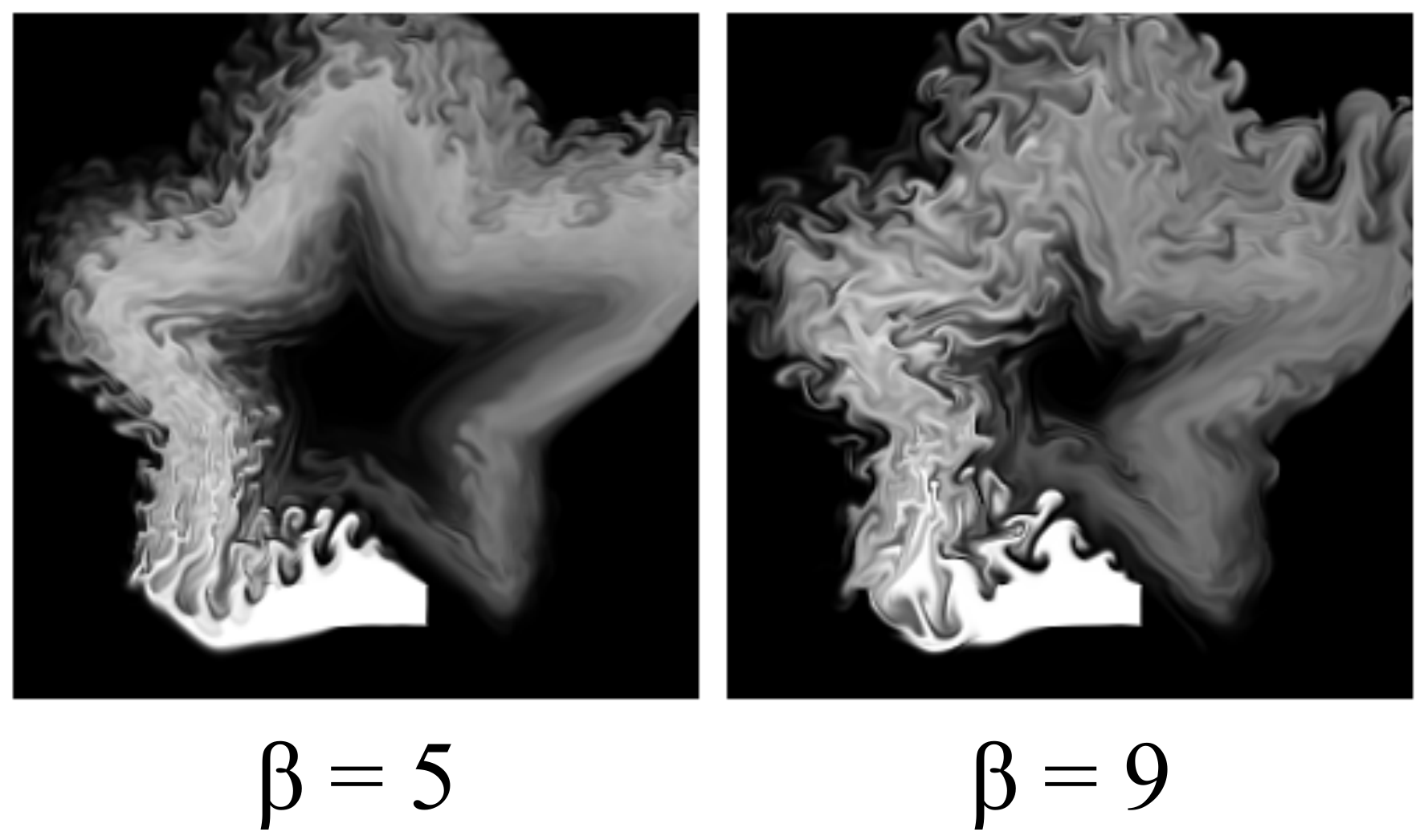}
		\caption{without up-res}
		\label{fig:starGuided}
	\end{subfigure}
	\vspace{-1mm}
	\caption{Simulations guided by a star-shaped target.
	}
\end{figure}

To realize interesting guided flows, we found an iterative {\em up-res} workflow that works well in practice. This is in line with previous work, where a low-resolution
input is refined in subsequent stages to yield a final result \cite{kim2008wavelet,Yuan2011,previewBasedSampling2011,flowFieldMod2013,Huang2013}.
Examples of this process are shown in \myreffig{fig:starGuided_upsample}. We start with a \mbox{$64^2$} simulation guided by a synthetic star-shaped velocity field with \mbox{$\SVW=1$} and \mbox{$\blurR=1$}. 
Then we run \mbox{$256^2$} high-resolution simulations with various $\blurR$ values. 
If the same guiding is done without up-res, that is, if we run the \mbox{$256^2$} guided simulation directly as shown in \myreffig{fig:starGuided}, 
then we lose the ability to guide the fluid to desirable shapes at multiple levels.

\begin{figure*}[tb!]
	\centering
	\includegraphics[width=0.93\textwidth]{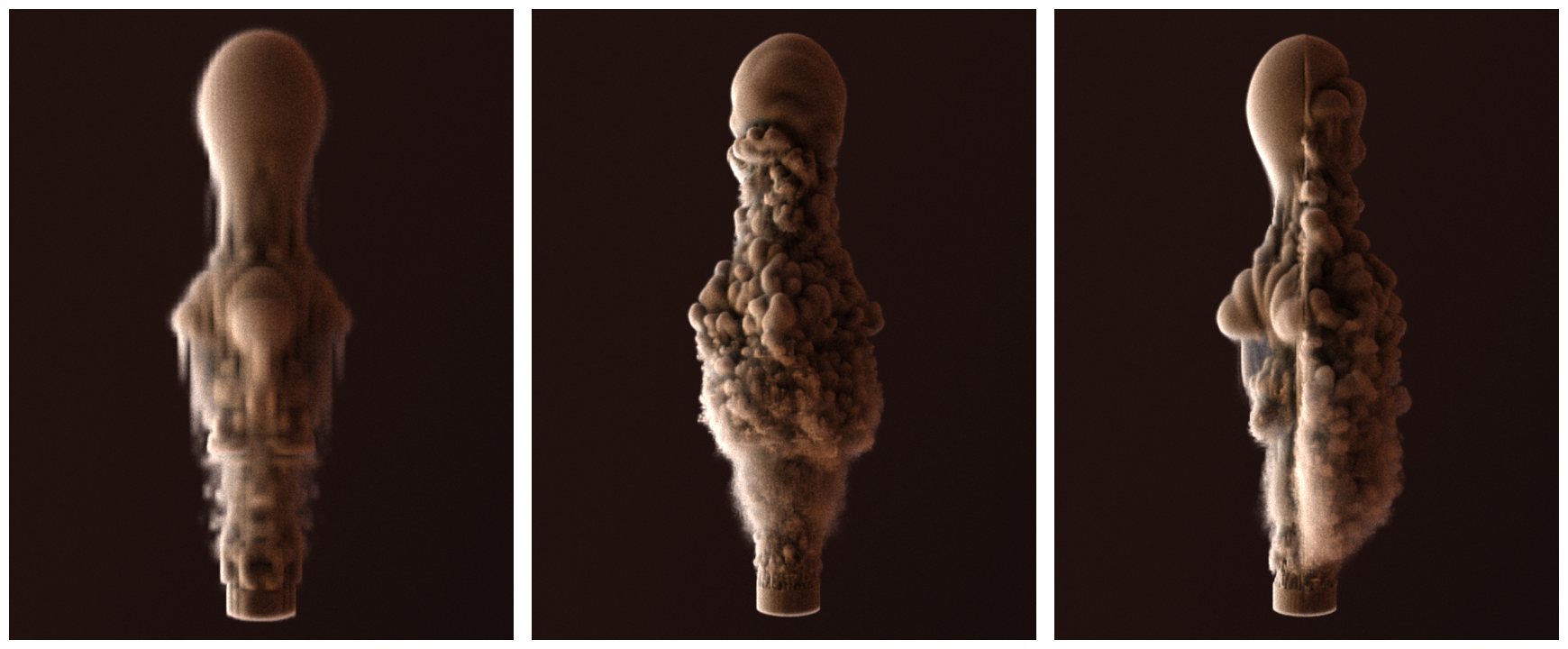}
	\vspace{-2mm}
	\caption{
		A \mbox{$50^2 \times 100$} simple plume simulation (left) is upsampled to \mbox{$200^2 \times 400$} with constant blur (middle, \mbox{$\blurR = 5$}) and spatially varying blur (right, \mbox{$\blurR = (1,10)$}).}
	\label{fig:simpleplume3D}
	\vspace{3mm}
	\centering
	\includegraphics[width=0.93\textwidth]{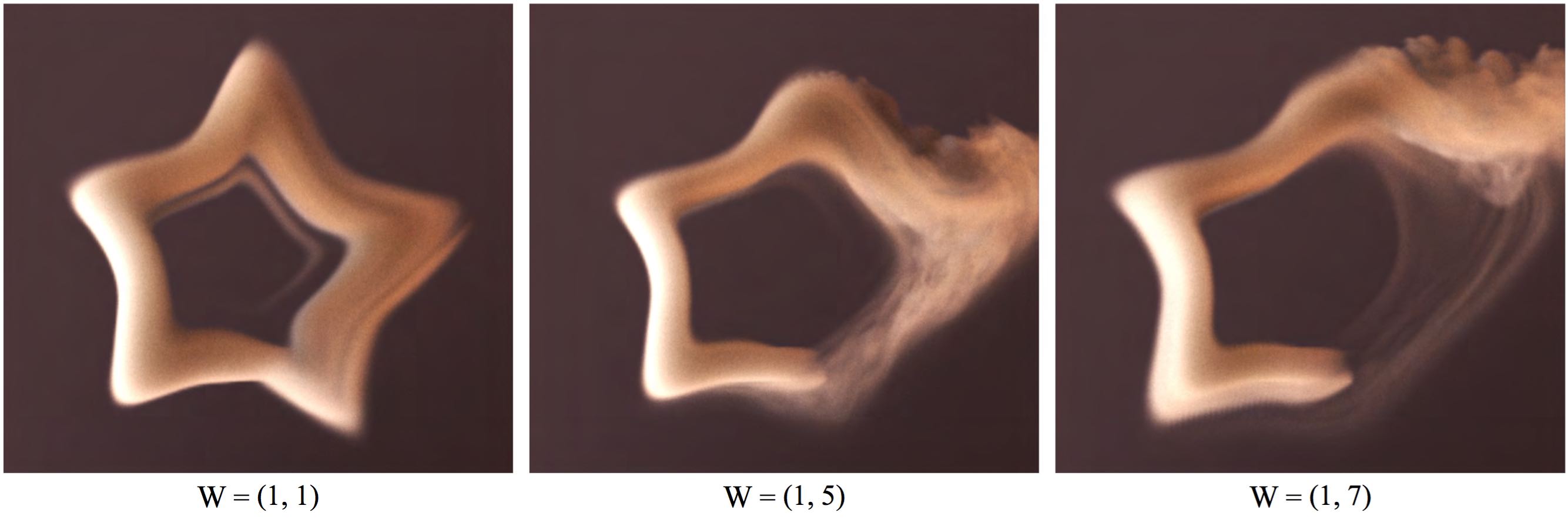}
	\vspace{-2mm}
	\caption{
		A \mbox{$250^3$} star-shaped guided simulation upsampled from a \mbox{$50^3$} guided simulation with spatially varying weights.}
	\label{fig:shapes3D}
	\vspace{3mm}
	\centering
	\includegraphics[width=0.93\textwidth]{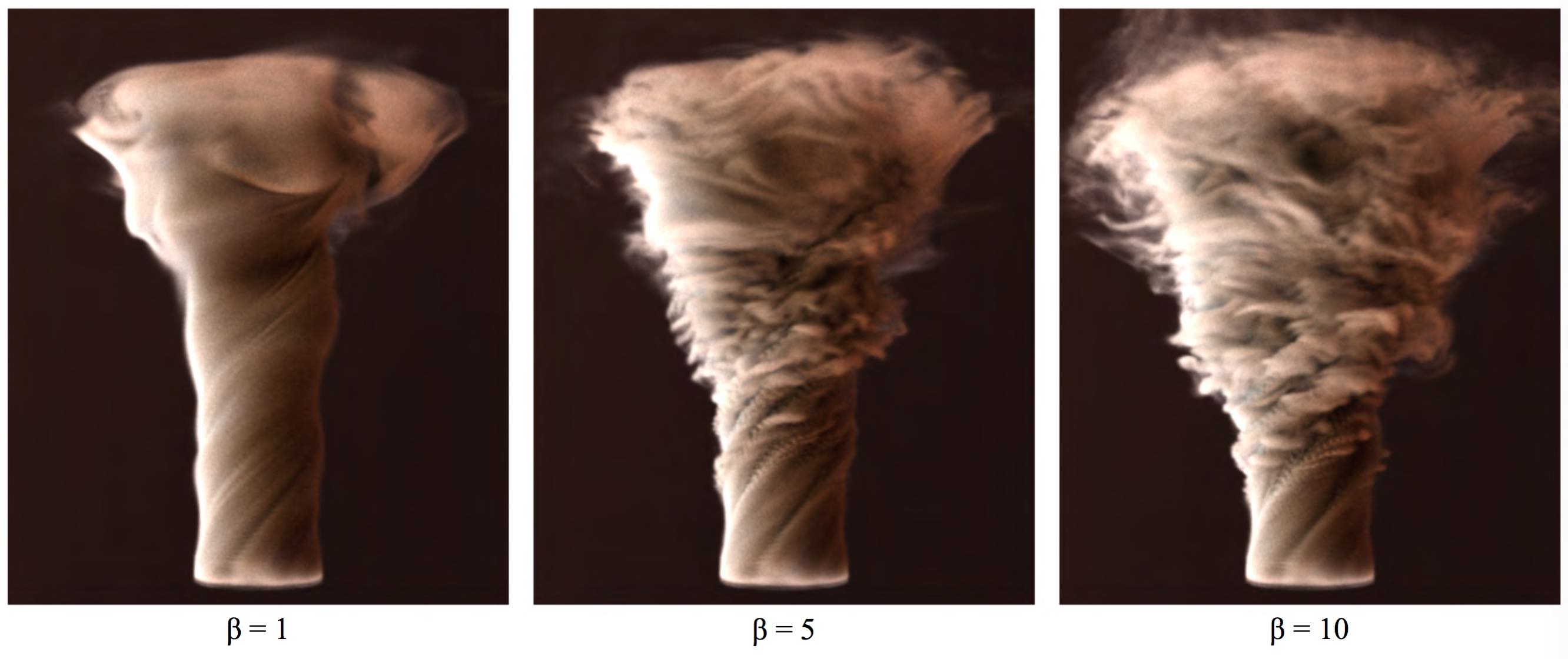}
	\vspace{-2mm}
	\caption{A \mbox{$200^2 \times 300$} guided simulation upsampled from a \mbox{$40^2 \times 60$} guided simulation with varying $\blurR$.}
	\label{fig:tornado}
\end{figure*}

We further demonstrate this up-res process for 3D simulations.
\myreffig{fig:simpleplume3D} shows a simple plume example. We first run a low-resolution (\mbox{$50^2 \times 100$}) simulation to capture the velocities. 
Then a \mbox{$200^2 \times 400$} resimulation is performed, guided by the upsampled velocity field. The blur radii can be kept spatially invariant or spatially varying for different effects. 
In particularly, increasing the blur radius introduces more small-scale details. 
In our spatially varying example, we use a blur kernel with a sharp transition between two blur radii to contrast between the two guided regions. 
If desired, this obvious seam can be softened via a more gradual transition between the two blur radii. 
Note that even though the simulation uses the approximation $\blurM^T \blurM = \blurM^2$, it works quite well in practice.

Our second example in \myreffig{fig:shapes3D} shows 3D simulations following star-shaped input velocities.
We use a $50^3$ target velocity as guide for a $250^3$ final resolution, with \mbox{$\blurR=2$}. 
Here we make use of spatially varying weights, with \mbox{$\SVW=1$} on the left side of the domain, and \mbox{$\SVW \in \{1,5,7\}$} on the right.
The precomputation time is negligible
compared to the overall run time: around $1.5$ seconds per time step for a $250^3$ simulation.

Next, we show a tornado simulation. We first run a low-resolution (\mbox{$40^2 \times 60$}) simulation using a cylindrical target velocity with a small upward component. 
Once the general shape is fixed, a \mbox{$200^2 \times 300$} resimulation is performed, guided by the upsampled velocity field. 
The results in \myreffig{fig:tornado} show the effectiveness of varying $\blurR$ to achieve different levels of turbulence.

\begin{figure*}[tb!]
	\centering
	\includegraphics[width=0.95\textwidth]{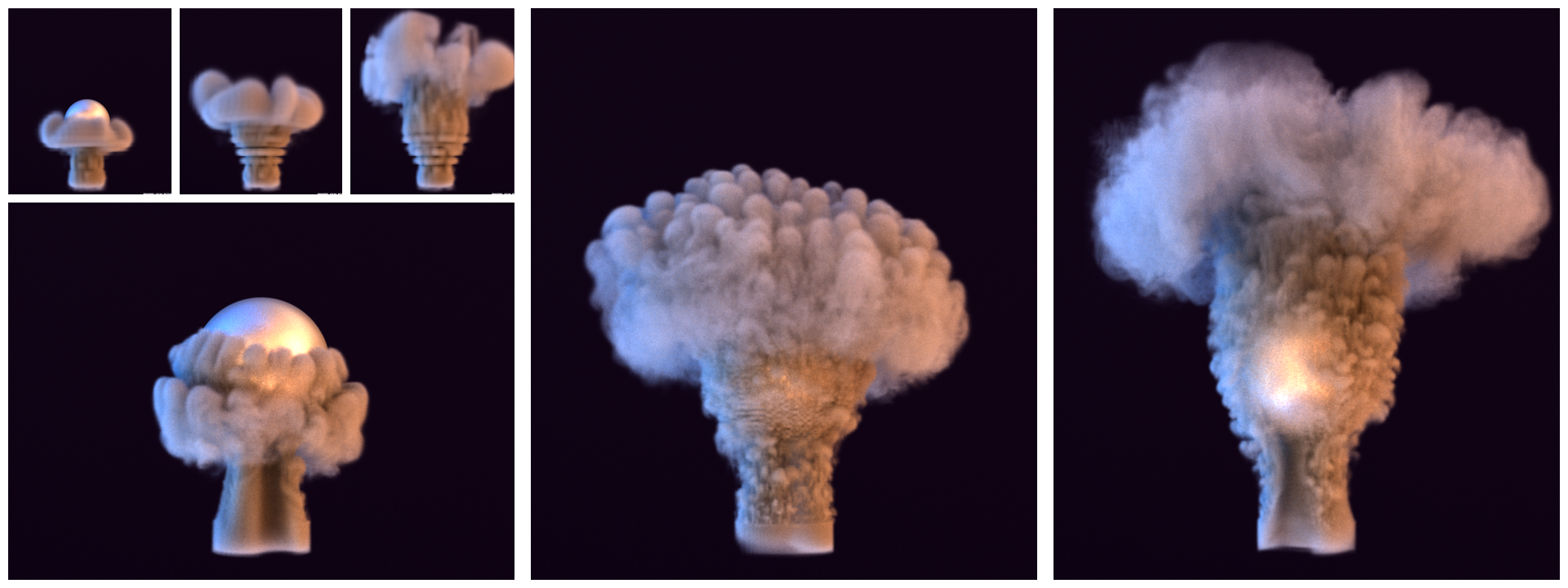}
	\caption{A \mbox{$40^3$} simulation with an obstacle (inset) is upsampled to \mbox{$200^3$} with $\SVW=1$ and $\blurR=5$. }
	\label{fig:obstacle3D}
\end{figure*}

\begin{figure}[tb!]
	\centering
	\includegraphics[width=0.48\textwidth]{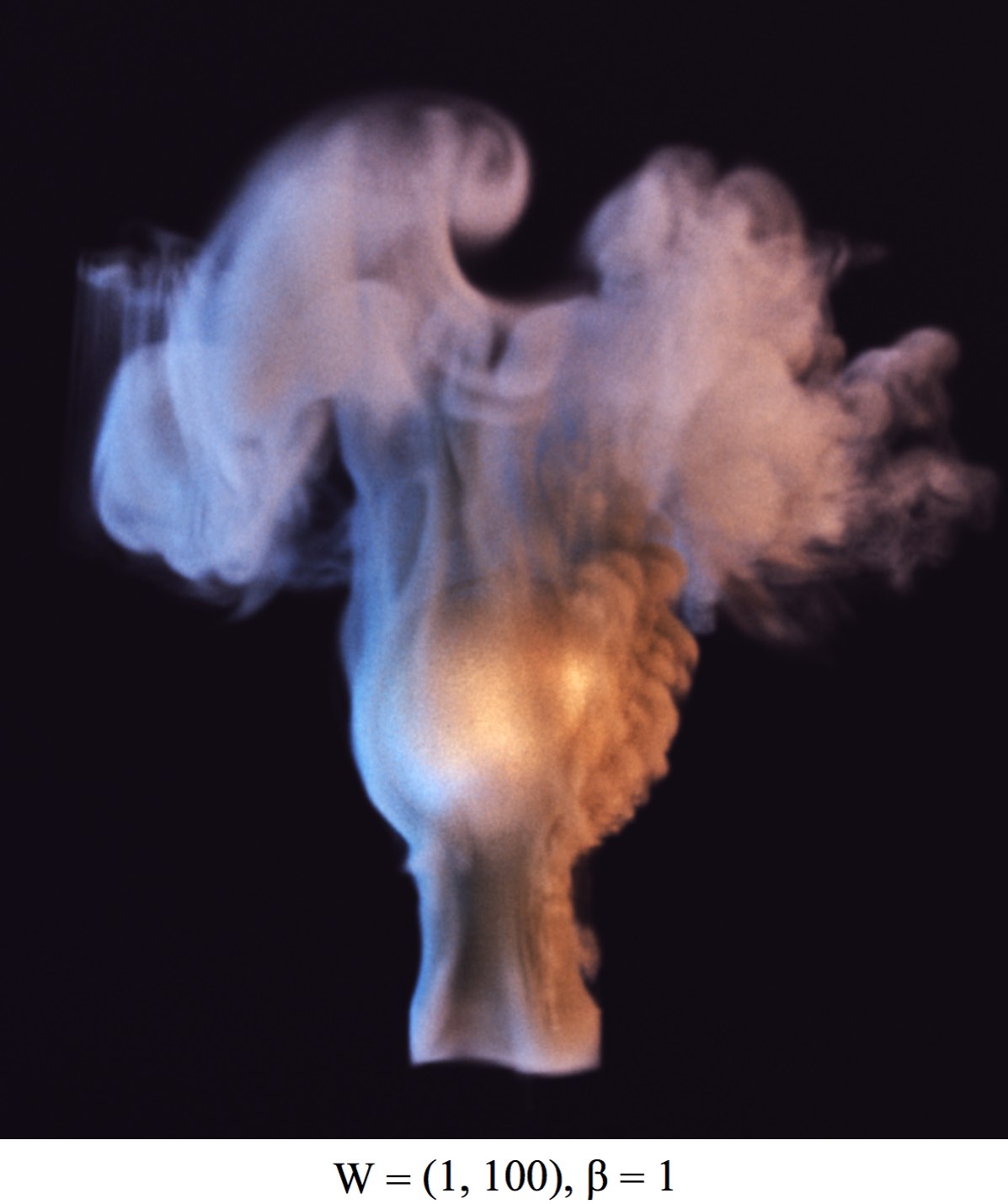}
	\vspace{-6mm}
	\caption{Guided obstacle scene with spatially varying weights.}
	\label{fig:obstacleSVW}
\end{figure}

Finally, \myreffig{fig:obstacle3D} demonstrates an example with obstacle, for which we upsampled from \mbox{$40^3$} to \mbox{$200^3$} with \mbox{$\SVW=1$} and \mbox{$\blurR=5$}. 
This high-resolution version captures the input motion, but develops many interesting small-scale details.
\myreffig{fig:obstacleSVW} shows a similar simulation with \mbox{$\blurR=1$} and spatially varying weights, where the left side has \mbox{$\SVW=1$} while the right side has \mbox{$\SVW=100$}. 
The result has little detail on the left side due to strong guiding from the low-resolution simulation, whereas the right side behaves like a regular smoke simulation due to weak guiding.
This guided simulation with an obstacle is a case that cannot be handled by previous Fourier-domain
guiding schemes~\cite{GregsonCapture} due to the interior boundaries. 
Arbitrary boundaries are easily handled by our separable approximation to the
fluid guiding proximal operator while remaining highly efficient and easily
parallelizable.

\section{Separating Solid-Wall BCs}
\label{sec:solidBCs}
\begin{figure}[tb!]
	\centering
	\includegraphics[width=0.47\textwidth]{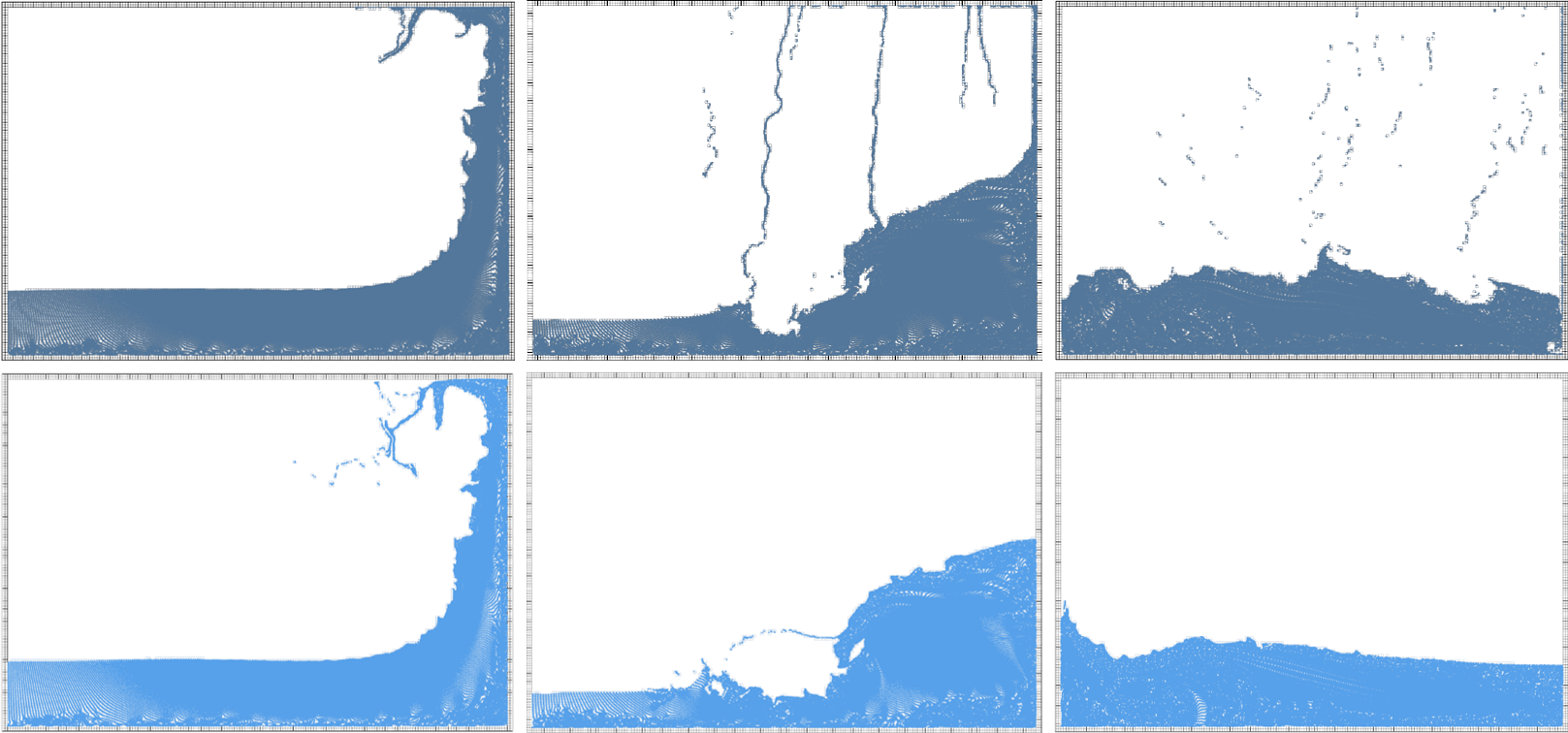}
	\caption{$200\times140$ Breaking Dam, Non-Separating (top) versus Separating BCs (bottom) at  $t=75, 112, 150$.}
	\label{fig:2DVisual}
\end{figure}

Another application of our PD-based method is the realization of separating solid-wall BCs.
As a motivational example, consider \myreffig{fig:2DVisual}, which shows a 2D breaking dam simulation. 
When generated by a common CG pressure solver (top row), the fluid exhibits the
undesirable behavior of sticking to the ceiling and getting stuck in corners.
In contrast, our method (bottom row) features a clean separation of the
fluid from all solids. In many visual effect settings that aim for large-scale fluid flow, the separation is preferable due to its improved realism.

We achieve this behavior by implementing separating solid-wall BCs in the form of
an inequality constraint \mbox{$\u\cdot\hat{n}\geq0$} 
(with $\u$ and $\hat{n}$ denoting velocity and obstacle normal)
without transforming the linear system of equations for the pressure solve into 
a more complicated problem.
As outlined in \myrefsec{sec:methodology}, we split the problem into two simpler objective functions, with $f$ controlling the BCs and $g$ enforcing zero divergence (see \myrefeq{eq:proxG}).
With this setup, a regular CG solver is employed to
compute $g$, while $f$ is handled by an efficient projection and
classification scheme, as described next.

\paragraph*{Velocity BCs}
The proximal operator 
\mbox{$\prox_{f, \sigma}(\gv)$} ensures that velocities at
obstacle surfaces never point into the solid (i.e., negative normal
velocity components at fluid-solid faces). Separating velocities  (i.e., positive normal components), on the other hand, are allowed and thus left alone.  We assume an obstacle is
static and has a velocity of zero at its surface.
BCs that form a linear constraint on the velocity field, like requiring velocity
components to be zero, can be expressed as orthogonal projections \cite{MolemakerLow}. 
Therefore, \mbox{$\prox_{f, \sigma}(\gv)$} reduces to a projection 
onto the space of velocities without flows into obstacles.
In its simplest form, this proximal operator sets normal velocity components to zero
wherever \mbox{$\u\cdot\hat{n}<0$}. 
The pressure solver, on the other hand, is unaware of any obstacle
boundaries and uses free surface Dirichlet BCs at liquid surfaces. 
Later, we present an accelerated version that makes the pressure solver partially aware of obstacles. 
For now, all obstacles are fully handled by $f$.

Although this simple velocity projection works for exact solutions, it breaks down when solving up to finite accuracy. The pressure solver 
can accumulate small positive velocities during iterations of our optimization procedure, 
leading to small separating motions where the liquid should only be standing still or moving tangentially.
This can happen for hydrostatic cases, for instance, where the final velocity at the wall should be exactly zero.
We address this problem by introducing a classification step with hysteresis.

We assume by default that our cells are separating boundary cells with a
Dirichlet $\Pre=0$ condition, and then create a list of cells that are explicitly not allowed to separate.
For the classification, in each iteration of our algorithm, we accumulate all motions into a wall
returned by the CG solver for a cell $i$ and store it in $m_i$.  This allows for a solid cell classification with temporal coherence. 
A non-separating cell only becomes separating if the magnitude of its velocity component
away from the wall $\u_i\cdot\hat{n}_i$ exceeds the magnitude of $m_i$.
Additionally, a change in the separation classification is only made if
the absolute value of $\u_i\cdot\hat{n}_i$ is greater than the accuracy of the CG solver
$\epsCG$. Otherwise, the cell keeps its previous state. 
The last two steps effectively implement a hysteresis that prevents cells
from changing status due to numerical errors, which is crucial for a
reliable convergence of our method.

\begin{figure}[tb!]
	\centering
	\includegraphics[width=0.44\textwidth]{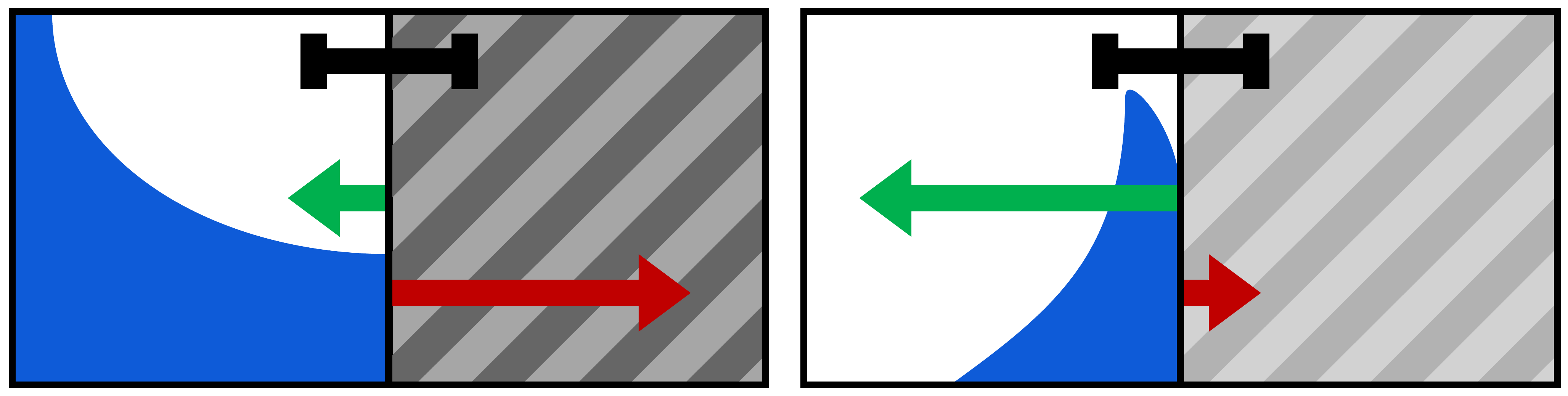}
	\caption{Classification of solid cells into non-separating (left) and separating (right). The green and red arrows denote $\u_i\cdot\hat{n}_i$ and $m_i$ respectively, while the black bars indicate $\epsCG$.
	}
	\label{fig:classification}
\end{figure}

\myreffig{fig:classification} illustrates this process.  It shows two possible
fluid behaviors at the face of a solid: non-separating (left) and separating
liquid (right).  Both velocities fulfill $|\u_i\cdot\hat{n}_i|>\epsCG$ (black bars), and
can thus potentially change state.
On the left of \myreffig{fig:classification}, the cell face has a large
accumulated value $m_i$ (red arrow). 
Assuming that the face velocity $\u_i\cdot\hat{n}_i$ (green arrow) exceeds the threshold
$\epsCG$, the cell is classified as non-separating, and $\u_i\cdot\hat{n}_i$  is later on set to zero despite
its positive value.
In contrast, a positive normal velocity component 
greater than $|m_i|$ indicates that fluid is currently separating from the solid
cell. The liquid boundary is treated as a free surface in this case, and its
velocity is left unmodified.

\paragraph*{Algorithm Summary}
\myrefalg{alg:proxFBC} shows pseudocode for the classification and the
proximal operator for separating BCs.  The classification proceeds as
outlined above, updating the list of non-separating cells $\FSn$ based on the velocity field $\u$.  The normal velocity components at non-separating cell faces
are set to zero in the proximal operator $\prox_{f, \sigma}(\gv)$. The variable $\gv$ is a generic argument variable being a combination of multiple velocity fields, including dual variables.

The PD algorithm solving for separating BCs follows the generic PD
implementation outlined in \myrefalg{alg:PDgeneric}.  We exchange the generic
proximal operator by our BC specific projection from \myrefalg{alg:proxFBC}.
Additionally, we call the classification function with $\z$ after line 6 of
\myrefalg{alg:PDgeneric}.

\begin{algorithm}[bt!]
	\caption{Classification and $\prox_{f,\sigma}$ for BCs}
	\label{alg:proxFBC}
	\begin{algorithmic}[1]
		\Procedure{classify}{$\u$, $m$, $\epsCG$, $\FSn$}
		\ForAll{boundary faces $i$, with $\left| \u_i \cdot \hat{n}_i\right| \geq \epsCG$}
		\If{$\u_i\cdot \hat{n}_i \leq 0$}
		\State $\FSn \leftarrow \FSn \cup \{i\}$ \ \ \ \text{// non-separating}
		\State $m_i \leftarrow m_i + \u_i\cdot \hat{n}_i$ 
		\ElsIf{( $|\u_i\cdot \hat{n}_i| \geq  |m_i|$ )}
		\State $\FSn \leftarrow \FSn \setminus \{i\}$ \ \ \ \text{// make separating}
		\State $m_i \leftarrow 0$ 
		\EndIf
		\EndFor
		\EndProcedure
		\Procedure{proxF}{$\gv$, $\FSn$}
		\State \text{// velocity projection for marked cells}
		\State \textbf{for all}\, $i \in \FSn$ \textbf{do}\, $\gv_i \leftarrow \gv_i-(\gv_i\cdot \hat{n})\hat{n}$ 
		\State \textbf{return}\, $\gv$
		\EndProcedure
	\end{algorithmic}
\end{algorithm}

We adopt two smaller modifications from previous work that
improve convergence.  The first is a Krylov method from IOP
\cite{MolemakerLow}, which we identified to work nicely within our BC solver
(the full algorithm can be found in \myrefapp{app:ADMMIOP}).  Additionally, we use the
adaptive parameter scheme~\cite{ChambollePD} for dynamic values of $\tau$,
$\sigma$ and $\theta$.  The parameter updates are the following after choosing
a suitable value for $\gamma$, $\tau^0$ and $\sigma^0$ (we use $\gamma=200$, $\tau^0=150$, $\sigma^0 = 1/\tau^0$):
$\theta^k \leftarrow 1 /\sqrt{1 + 2\tau^{k-1}\gamma}$, $\tau^k \leftarrow \tau^{k-1}\theta^k$, and $\sigma^k \leftarrow \sigma^{k-1} / \theta^k$.

Both methods reduce the overall run time, but are specific extensions for 
proximal operators given by orthogonal projections. As such, we use the extensions
for all BC problems, but not for our guided simulations.

\paragraph*{Accelerated BC Solver}
So far, 
the solid-wall BC handling is fully done in $f$ while $g$ simply ensures incompressibility.
This is in line with the standard splitting approach in other methods \cite{MolemakerLow,Henderson2012}, 
and we demonstrate in the next section that our optimization scheme very efficiently
calculates the solution in this case.

However, in practice, the total number of iterations can be reduced
significantly by letting the pressure solver take care of all BCs that it is
capable of enforcing correctly: retaining the input normal velocity components at fluid-solid faces by enforcing a zero pressure gradient at walls.
For this version of our solver, we update the BCs of the
pressure solver in $g$ in accordance with our classification: Neumann BCs for
non-separating cells, and Dirichlet (free surface) BCs for
separating ones. 
This effectively locks the classification after few iterations, leading to a stable solution within the next iteration. 
However, it prevents non-separating cells from changing their state back to separating ones. Thus, the
accelerated BC solver does not yield the same result as the standard version, but
achieves much higher performance while featuring only negligible differences. A summary of our accelerated solver with pseudo-code can be found in \myrefapp{app:accBCSolver}.

\subsection{Evaluation and Results}
\label{sec:resultsBC}
We now evaluate the performance of our BC solver, and demonstrate
the importance of allowing liquid to separate from walls in high-resolution simulations\punctuationfootnote{
	Thresholds for our algorithm are set to \mbox{$\epsCG=10^{-5}$} (as liquids are more sensitive to mass loss) and \mbox{$\epsAbs=\epsRel=10^{-3}$}.
}.
To validate that our PD solver yields the correct results, we use our
method to calculate regular non-separating boundaries
for a $200\times140$ breaking dam setup, which can also be handled by a
regular CG solver. In this case, we can achieve arbitrary accuracies depending
on the choice of parameters for CG and our scheme. These
experiments show that our method converges to the correct solution.

\begin{figure*}[tb!]
	\centering
	\begin{subfigure}{0.33\textwidth}
		\includegraphics[width=\textwidth]{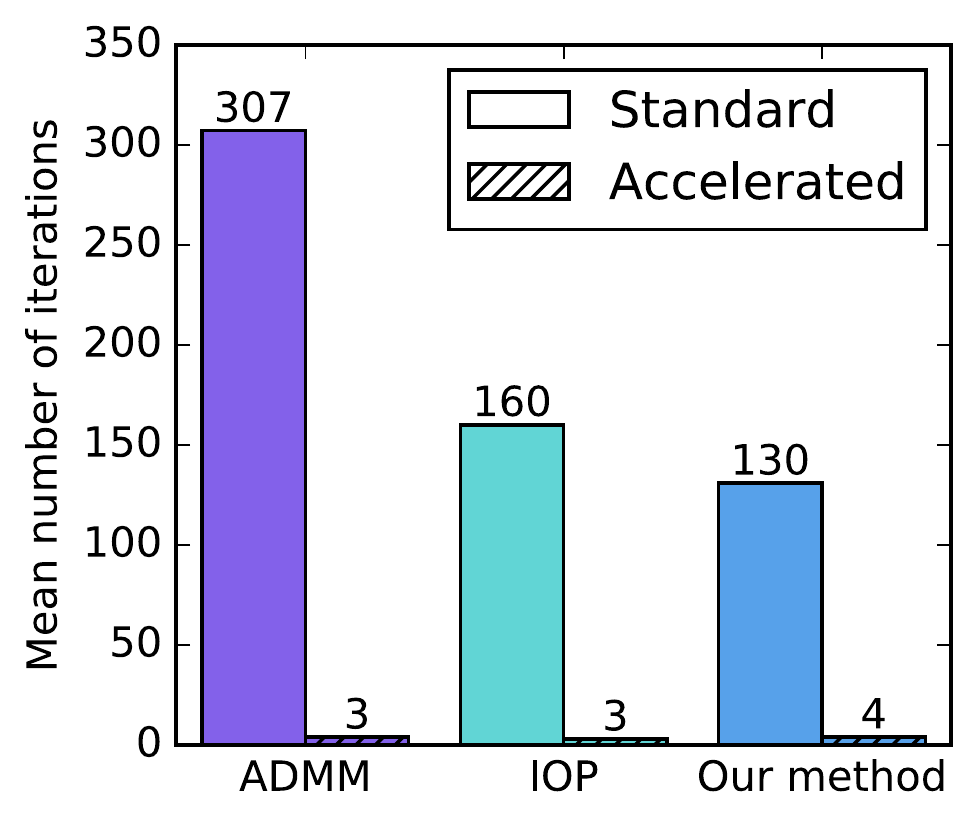}
		\caption{IOP/ADMM/PD Iter}
		\label{fig:2DIter}
	\end{subfigure}
	\begin{subfigure}{0.33\textwidth}
		\centering
		\includegraphics[width=\textwidth]{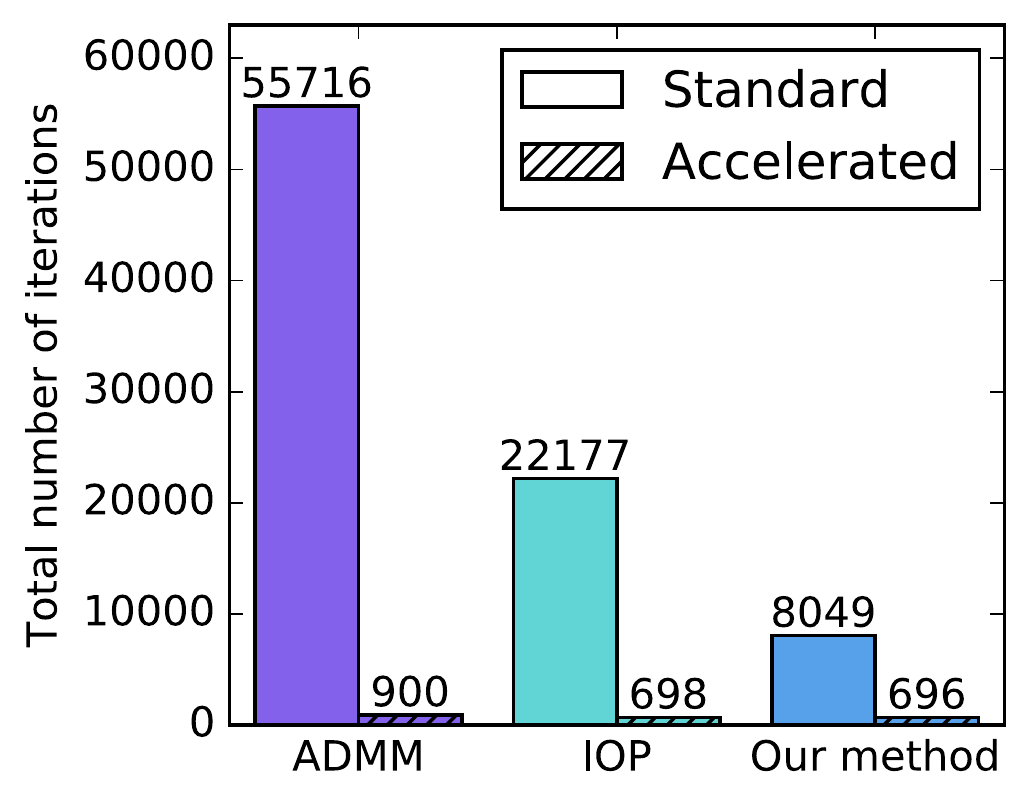}
		\caption{Sum of CG Iter}
		\label{fig:2DSIter}
	\end{subfigure}
	\begin{subfigure}{0.33\textwidth}
		\centering
		\includegraphics[width=\textwidth]{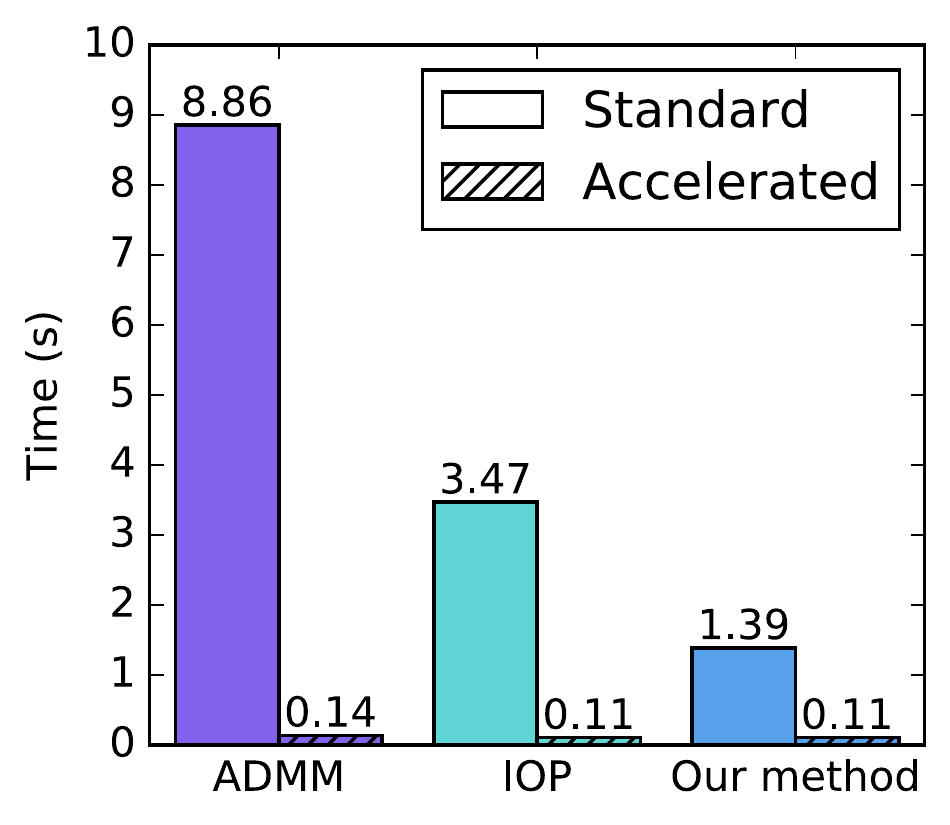}
		\caption{Runtime in seconds}
		\label{fig:2DTime}
	\end{subfigure}
	\caption{2D breaking dam, mean quantities per time step.}
	\label{fig:2DBC}
\end{figure*}

\begin{figure}[tb!]
	\centering
	\includegraphics[width=0.33\textwidth]{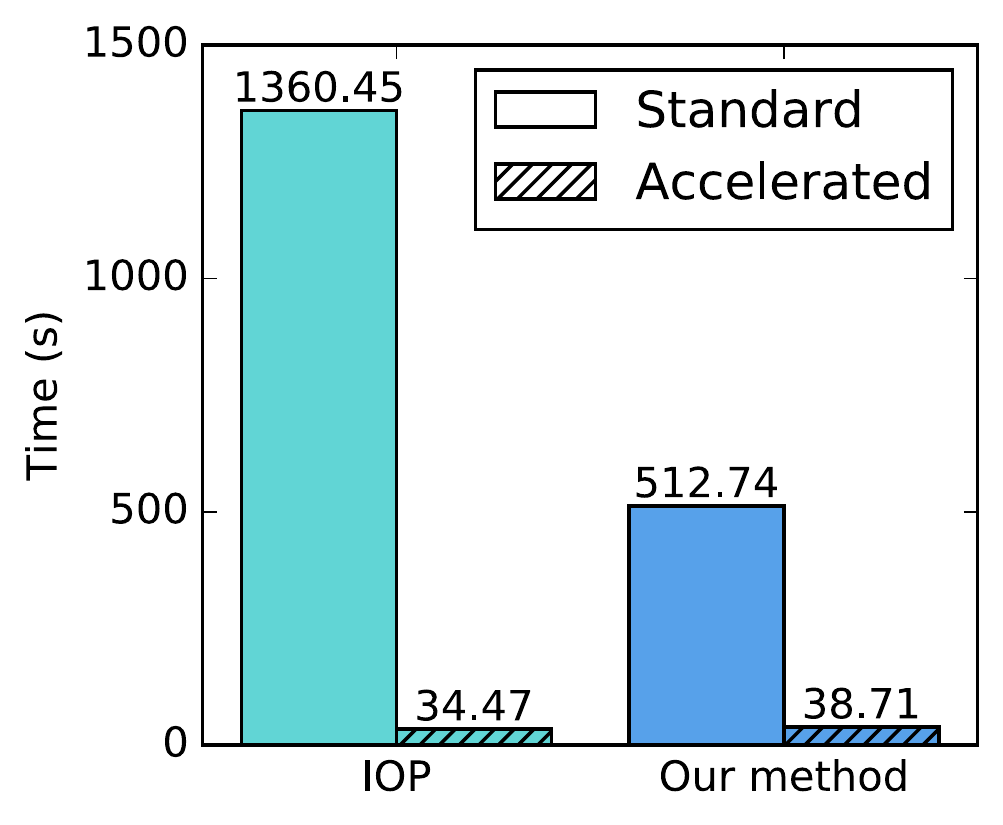}
	\caption{3D complex breaking dam, mean run time per time step.}
	\label{fig:BCPerformance}
\end{figure}

\begin{figure*}[tb!]
	\centering
	\includegraphics[width=0.95 \textwidth]{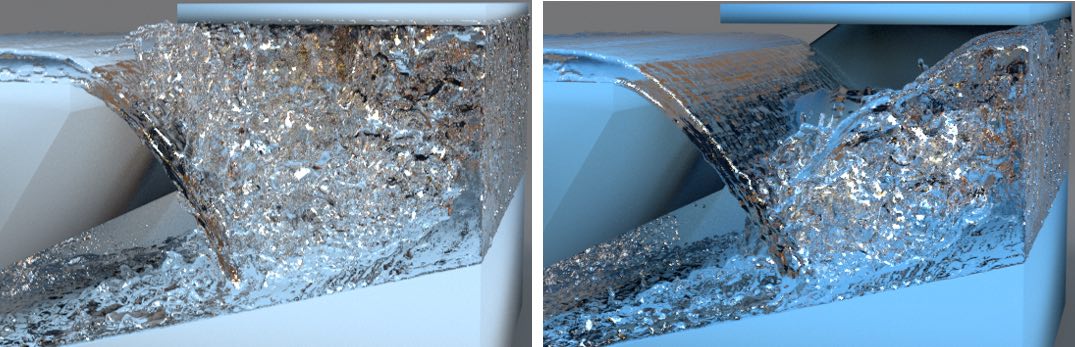}
	\caption{ A \mbox{$256\times230\times256$} scene with complex obstacles. Regular, non- 
		BCs lead to the liquid sticking to walls (left), while our solver allows for naturally separating	liquids (right). }
	\label{fig:visual3D}
\end{figure*}

For separating boundaries, we can no longer compare to a regular CG solver. 
Instead, we evaluate the performance of our method against methods from previous convex optimization work: IOP and ADMM. 
IOP has been applied to enforce non-divergence and complex BCs simultaneously in \cite{ChentMG}. 
For the standard BC solver simulating a 2D breaking dam, our method converges six times faster than ADMM, and more than twice as fast as IOP, as shown in \myreffig{fig:2DTime}. 
\myreffig{fig:2DIter} and \myreffig{fig:2DSIter} show the mean number of IOP/ADMM/PD and CG iterations. 
Our method generally requires a lower number of PD iterations compared to ADMM and IOP. 
Due to our adaptive CG accuracy scheme, the number of PD iterations is increased while the total amount of CG iterations is decreased. 
Both numbers influence the performance, but the total number of CG iterations more strongly influences runtime as shown in \myreffig{fig:2DTime}.
Similar behavior is observed for a 3D complex breaking dam simulation (see \myreffig{fig:visual3D} for a visual example). 
The run time measurements are shown in \myreffig{fig:BCPerformance} where ADMM is omitted due to its impractical run time.
The accelerated solver speeds up all methods significantly. In this case, our PD-based method
performs on par with IOP, which we attribute to the low number of iterations required in this case;
the higher-order convergence of the PD method does not develop its full potential in such a setting.

\myreffig{fig:visual3D} highlights that regular non-separating walls often yield undesirable results
in practical settings. We simulate a complex 3D scene with liquid splashing onto multiple
obstacles with a resolution of $256\times230\times256$. A regular solver leads
to large amounts of liquid crawling along ceilings, and sticking to walls,
while our separating boundaries give a much more believable large scale look,
with liquid naturally separating from obstacle boundaries.
For this setting, our accelerated BC solver requires $38.7s$ on average per
frame. Comparing the overall time to generate this result (including surface
generation) to a version with a regular pressure solver, our method increases
run time by only $12\%$, with the added benefit of enabling separating BCs. 
This is a very practical result, considering that it was achieved based on a
regular CG solver, without the need for specialized methods, such as Linear Complementarity Problems
solvers \cite{gerszewski2013physics}. 
Although solving LCPs has been studied in detail, these methods still fall into complexity classes that make large scale solves infeasible. An indication for this can be found in \cite{gerszewski2013physics}, where solving times of ca. 25s per LCP solve were given for a $100^3$ example. 

\section{Discussion and Conclusion}
We presented a general framework for incorporating the Primal-Dual method into
fluid simulation, and demonstrated two applications: fluid guiding and 
separating BCs.
We proposed a generic version of the Primal-Dual based optimization scheme
with fast convergence for general proximal operator subproblems. In addition,
we discussed several extensions that are particularly well-suited for optimizations
with orthogonal projections as subproblems.
Additionally, we demonstrated a novel formulation for the flow guiding
problem, and an efficient approach for simulating liquids with separating 
BCs.

\paragraph*{Limitations}
One limitation of our fluid guiding method is a lack of shape controls. Unlike
smoke, liquids can require shape constraints to achieve a desired outline or
shape.  We have focused on velocity guiding in this work, so controlling the
shape of liquids will require extensions that take the position of
the liquid's surface into account. Such constraints should integrate well into our overall
pipeline, and we plan to investigate this topic as future work.
Another area of improvement is bridging the gap between our optimization framework and an artist's workflow. 
Although we gave examples of different target velocities used for guiding, there is still a lot more to explore in terms of how to achieve various artistic visual effects intuitively.
A limitation of our accelerated separating BC solver is that it can lead to slight deviations from the
accurate, standard solution. We have not encountered visual artifacts resulting from these inaccuracies, and we believe that the
improved performance typically outweighs these slight deviations.
The specialized multigrid solvers \cite{ChentMG} could possibly outperform our BC solver. 
We believe that the attractiveness of our BC solver stems from its modularity and fast convergence. 
Adding support for wall separating BCs given an existing pressure solver requires little code with our method.

\paragraph*{Outlook}
Our method is a very generic approach applicable to a large range of problems in fluid simulation.
The price of this generality is that it may not be as fast as specialized methods tailored to specific problems.
However, the modularity of our approach makes it easy to incorporate into existing implementations.
As such, it has the potential to add powerful functionality, such as high-level
flow guiding, into existing solvers without the need for complicated
extensions.

Furthermore, 
there is a large number of interesting avenues
to be explored with high-level optimizations of fluids flows.
For example, we are interested in exploring shape optimization to adapt
the geometry of obstacles with respect to their flow properties,
and partial resimulations of regions in a flow.

\section{Acknowledgments}
This work was funded by the ERC Starting Grant {\em realFlow} (StG-2015-637014) and the NSERC Postdoctoral Fellowships Program.

\appendix

\small

\vspace{-0.2cm}
\section{Notation}
\label{app:notation}

For reference we provide a quick description of our notation. 
More detailed descriptions can be found in the body of the paper.
\begin{itemize}
	\item $f$: first objective function in PD (application-dependent)
	\item $g$: second objective function in PD for incompressibility
	\item $\gv$: general input velocity field for several algorithms, a combination of several velocity fields, possibly dual variables
	\item $\x$, $\z$, $\y$: variables with iterative updates in PD
	\item $\tau$, $\sigma$, $\theta$: PD parameters controlling convergence rate
	\item $\alpha$, $\blurR$, $\blurM$, $\q$, $M$, $\ut$, $\uc$, $L$: variables specific to fluid guiding
	\item $\u$: fluid velocity for the BC problem
	\item $\hat{n}$: normal of a solid cell
	\item $m$: memory velocity field
	\item $\FSn$: set of non-separating fluid-solid faces
\end{itemize}

\vspace{-0.2cm}
\section{ADMM and IOP}
\label{app:ADMMIOP}

Iterated Orthogonal Projection (IOP)~\cite{MolemakerLow}---a method similar to von Neumann's alternating projections~\cite{Boyd:ADMM}---requires both subproblems to be expressed as orthogonal projections
\begin{align}
\x^{k+1} &= \PiB_{f}(\z^k)\\
\z^{k+1} &= \PiB_{g}(\x^{k+1}).
\end{align}
The Krylov method can improve its convergence rate:
\begin{algorithmic}[1]
	\Procedure{krylov}{$\z^k$, $\z^{k-1}$, $k$, $\epsilon^{k-1}$}
	\State $\epsilon^k = \text{error}(\z^k)$
	\If {$k \textgreater 1$}
	\State $\z_{\dif} = \z^k - \z^{k-1}$ // correction vector
	\State $\epsilon_{\mathrm{ratio}} = \epsilon^k / \epsilon^{k-1}$ 
	\State $\z_{\tmp} = \z^k - \epsilon_{\mathrm{ratio}} \z_{\dif}$ 
	\State $\epsilon_{\tmp} =\text{error}(\z_{\tmp}) $ 
	\State \textbf{if}\, {$\epsilon^{\tmp} \textless \epsilon^k$} \textbf{then }$\z^k = \z_{\tmp}$ 
	\EndIf
	\EndProcedure
\end{algorithmic}

The Alternating Direction Method of Multipliers (ADMM) \cite{Boyd:ADMM,GoldsteinADMMPD} is a proximal method  more general than IOP given by
\begin{align}
\x^{k+1} &:= \prox_{f,\rho}(\z^k - \y^k) \\
\z^{k+1} &:= \prox_{g,\rho}(\x^{k+1} + \y^k) \\
\y^{k+1} &:= \y^k + \x^{k+1} - \z^{k+1}.
\end{align}
Instead of the three parameters $\{\tau,\sigma,\theta\}$ in PD that control convergence rate, 
ADMM only has one such parameter, $\rho$.

\vspace{-0.2cm}
\section{Inverse Matrix Approximation for Fluid Guiding}
\label{app:invM_derivation}

Here we present the details for deriving the inverse matrix approximation in \myrefeq{eq:invM_approx}. 

We want to invert a matrix of the form $M = A + (2\blurM^T \blurM)$, where $A = 2 \SVW^2 + \sigma I$ contains the large diagonal terms and $2\blurM^T \blurM$ the small off-diagonal terms. By the Sherman-Morrison-Woodbury Formula,
\begin{align}
M^{-1} &= A^{-1} - 2 A^{-1} \blurM^T (I + 2 \blurM A^{-1} \blurM^T) \blurM A^{-1}.
\end{align}
Now since $\blurM$ and $A^{-1}$ both contains small value entries, $2 \blurM A^{-1} \blurM^T$ is approximately zero. Hence
\begin{align}
M^{-1} &\approx A^{-1} - 2 A^{-1} \blurM^T \blurM A^{-1} \\
&= (2 \SVW^2 + \sigma I)^{-1} - 2 (2 \SVW^2 + \sigma I)^{-1} \blurM^T \blurM (2 \SVW^2 + \sigma I)^{-1}.
\end{align}
Note that the calculation of $A^{-1}$ is trivial since $A = 2 \SVW^2 + \sigma I$ is diagonal.

\section{Accelerated BC Solver}
\label{app:accBCSolver}
Below, we summarize our accelerated solver for separating solid-wall BCs. 
Instead of fully handling the solid-wall BCs in a proximal operator outside of the pressure projection, our accelerated
solver employs the commonly used Neumann BCs during the pressure projection depending on the current classification of solid-wall cells.
First, all solid wall cells are classified as separating (i.e. Dirichlet BCs for the pressure solver).
Depending on the initial velocity field, solid wall cells with \mbox{$\u\cdot\hat{n}<0$} are classified 
as non-separating cells (Neumann BCs for the pressure solver).
We then iteratively set the velocity BCs, apply a regular CG pressure solver and re-classify solid cells with the \textproc{classify} procedure. 

In our standard BC solver (\myrefsec{sec:solidBCs}), a memory field $m$ is used to determine whether a cell is allowed to change its state to separating. 
In the accelerated case, $m$ is not used, instead the pressure solver enforces solid wall BCs for non-separating cells. This prevents 
the normal velocity from becoming positive, and from changing its state back to separating. 
In the following procedure, $m$ is replaced with the placeholder "$\cdot$".
\begin{algorithmic}[1]
	\Procedure{solvePressureWithSeparatingBCs}{$\u$, $\epsilon$}
	\State $\FSn={0}$ // no cell is classified as non-separating yet
	\State \textproc{classify}($\u$, $\cdot$, $\epsilon$, $\FSn$)
	\While{$\FSn$ did not change in last call of \textproc{classify}}
	\State $\textproc{proxF}_{BC}$($\u$, $\FSn$)
	\State $\PiDiv$($\u$, $\epsilon$)
	\State \textproc{classify}($\u$, $\cdot$, $\epsilon$, $\FSn$)
	\EndWhile
	\EndProcedure
\end{algorithmic}

\normalsize
\vspace{-0.2cm}


\renewcommand{\baselinestretch}{0.96} 
\bibliographystyle{eg-alpha}
\bibliography{CG_and_CV_conferences_and_journals,references}

\end{document}